\PassOptionsToPackage{unicode}{hyperref}
\PassOptionsToPackage{hyphens}{url}
\documentclass[
]{article}
\usepackage{amsmath,amssymb}
\usepackage{iftex}
\ifPDFTeX
  \usepackage[T1]{fontenc}
  \usepackage[utf8]{inputenc}
  \usepackage{textcomp} % provide euro and other symbols
\else % if luatex or xetex
  \usepackage{unicode-math} % this also loads fontspec
  \defaultfontfeatures{Scale=MatchLowercase}
  \defaultfontfeatures[\rmfamily]{Ligatures=TeX,Scale=1}
\fi
\usepackage{mathptmx}
\ifPDFTeX\else
\fi
\IfFileExists{upquote.sty}{\usepackage{upquote}}{}
\IfFileExists{microtype.sty}{% use microtype if available
  \usepackage[]{microtype}
  \UseMicrotypeSet[protrusion]{basicmath} % disable protrusion for tt fonts
}{}
\makeatletter
\@ifundefined{KOMAClassName}{% if non-KOMA class
  \IfFileExists{parskip.sty}{%
    \usepackage{parskip}
  }{% else
    \setlength{\parindent}{0pt}
    \setlength{\parskip}{6pt plus 2pt minus 1pt}}
}{% if KOMA class
  \KOMAoptions{parskip=half}}
\makeatother
\usepackage{xcolor}
\usepackage{longtable,booktabs,array}
\usepackage{calc} % for calculating minipage widths
\usepackage{etoolbox}
\makeatletter
\patchcmd\longtable{\par}{\if@noskipsec\mbox{}\fi\par}{}{}
\makeatother
\IfFileExists{footnotehyper.sty}{\usepackage{footnotehyper}}{\usepackage{footnote}}
\makesavenoteenv{longtable}
\ifLuaTeX
  \usepackage{selnolig}  % disable illegal ligatures
\fi
\IfFileExists{bookmark.sty}{\usepackage{bookmark}}{\usepackage{hyperref}}
\IfFileExists{xurl.sty}{\usepackage{xurl}}{} % add URL line breaks if available
\hypersetup{
  hidelinks,
  pdfcreator={LaTeX via pandoc}}

\author{}
\date{}

\usepackage{newunicodechar}
\newunicodechar{−}{\ensuremath{-}}
\newunicodechar{β}{\ensuremath{\beta}}
\newunicodechar{ᵢ}{\ensuremath{_{i}}}
\newunicodechar{→}{\ensuremath{\rightarrow}}
\newunicodechar{²}{\ensuremath{^{2}}}
\newunicodechar{ö}{{\"o}}
\newunicodechar{°}{\textdegree{}}
\newunicodechar{₂}{\ensuremath{_{2}}}
\newunicodechar{χ}{\ensuremath{\chi}}
\newunicodechar{·}{\textperiodcentered{}}
\newunicodechar{μ}{\ensuremath{\mu}}
\newunicodechar{σ}{\ensuremath{\sigma}}
\newunicodechar{ł}{{\l}}
\newunicodechar{ó}{{\'o}}
\newunicodechar{×}{\ensuremath{\times}}
\newunicodechar{α}{\ensuremath{\alpha}}
\newunicodechar{₁}{\ensuremath{_{1}}}
\newunicodechar{₃}{\ensuremath{_{3}}}
\newunicodechar{₄}{\ensuremath{_{4}}}
\newunicodechar{Σ}{\ensuremath{\Sigma}}
\newunicodechar{γ}{\ensuremath{\gamma}}
\newunicodechar{ₛ}{\ensuremath{_{s}}}
\newunicodechar{ε}{\ensuremath{\varepsilon}}
\newunicodechar{®}{\textregistered{}}
\usepackage[margin=1in]{geometry}
\usepackage{graphicx}
\usepackage{tikz}
\usetikzlibrary{arrows.meta,positioning,shapes.geometric}
\usepackage{float}
\usepackage{placeins}
\usepackage{needspace}
\usepackage{url}

\begin{document}

\title{Same Firms, Different Verdicts: ESG Rating Choice and the Measurement of Greenwashing}
\author{Praveen Kumar Ashok Kumar\\ \small QFAI Institute, KUE \and Rafa\l{} Sieradzki\\ \small NYU Stern, QFAI Institute, KUE, QFRG}
\date{}
\maketitle
\thispagestyle{empty}
\begin{center}
This version {\today} \\
\vspace{0.5cm}
\small\textbf{Corresponding author:} \\
\vspace{0.2cm}
 rafal.j.sieradzki@gmail.com
\end{center}
\vspace{1.5cm}
\begin{abstract}
\noindent This paper investigates the Aggregate Confusion hypothesis (Berg, K\"olbel, and Rigobon, 2022) at the firm level by measuring the Disclosure-Performance Gap (DPG), defined as the standardised divergence between a firm's voluntary environmental disclosure (``Talk'') and its realised emissions performance (``Walk''). The sample comprises 200 large European firms drawn from the Energy, Materials, Industrials, and Utilities sectors of the STOXX Europe 600 in fiscal year 2023, the final cross-section of the voluntary reporting era before the Corporate Sustainability Reporting Directive. The estimated model is identified through a systematic six-stage model selection process, namely candidate assembly, Pearson correlation screening, multicollinearity filtering by Variance Inflation Factor, stepwise forward search across five variable pools under the corrected Akaike Information Criterion, Cook's distance influence screening, and HC3 re-estimation, which evaluates 421 candidate specifications. The selected model is estimated by ordinary least squares with HC3 robust standard errors on the full sample of 200 firms. The strongest predictor of a wider gap is flagship index membership ($\beta = +0.78$, $p < 0.01$), consistent with institutional ceremonial conformity; Task Force on Climate-related Financial Disclosures (TCFD) endorsement is also positive ($\beta = +0.86$, $p < 0.05$) but is identified off a small group of non-supporting firms and is read as directional rather than as a precise magnitude. Two substantive commitments significantly narrow the gap, renewable energy use ($\beta = -0.31$, $p < 0.01$) and environmental capital expenditure ($\beta = -0.22$, $p < 0.05$), consistent with signalling theory. Governance and monitoring variables carry no explanatory power. The core findings are stable under influence trimming, under a rank-based recoding of the ordinal disclosure score, and when the TCFD variable is removed. Replacing the CDP Climate Score with the LSEG Environmental Pillar Score on the identical firms eliminates the index-membership effect while the renewable-energy effect survives, showing that detected greenwashing is conditional on the rating lens applied.
\end{abstract}

\vspace{4.5cm}

\begin{flushleft}

\end{flushleft}

\newpage

\section*{1.\quad INTRODUCTION}

The integration of Environmental, Social, and Governance (ESG) criteria into capital allocation is one of the most consequential structural changes in modern finance. ESG-related assets under management have surpassed USD 30 trillion and are projected to approach USD 40 trillion by 2030 (Bloomberg Intelligence, 2024). This accumulation of capital has outpaced the development of the informational infrastructure needed to direct it, producing a market characterised by data fragmentation and a credibility problem.

At the centre of this problem is the phenomenon of ``Aggregate Confusion'', the systematic divergence in environmental ratings across ESG assessment agencies first documented by Berg, K\"olbel, and Rigobon (2022). Unlike credit ratings, which exhibit inter-agency correlations approaching 0.99 (Becker \& Milbourn, 2011), ESG ratings frequently fall below 0.60. The inconsistency is not merely technical; it has allocative consequences. Investors who use ESG scores to identify climate transition readiness cannot reliably separate firms that are decarbonising from firms that have mastered the production of legitimacy through disclosure. The informational asymmetry between manager and investor, the classic agency problem of capital market theory, is amplified rather than resolved by the frameworks designed to address it.

The existing literature has largely framed greenwashing as a binary, ex post classification: firms are identified as non-compliant after litigation, regulatory censure, or NGO exposure. This approach carries two deficiencies. First, it suffers from survivorship bias, because scandal-based detection overlooks sophisticated policy-practice decoupling that remains sub-legal and undetected. Second, it is reactive, intervening only after capital has been misallocated. The gap between corporate ``Talk'' and corporate ``Walk'' is therefore likely to be underestimated by models that capture only the tail of the distribution.

This paper departs from that tradition. Building on the continuous conceptualisation of greenwashing developed by Delmas and Burbano (2011), the Disclosure-Performance Gap (DPG) is operationalised as a quantifiable spectrum: the standardised distance between a firm's voluntary CDP Climate Score, the primary institutional proxy for disclosure quality, and its Scope 1+2 emission intensity, the primary operational proxy for decarbonisation progress. Treating the gap as a continuous dependent variable allows detection of systemic decoupling across the full distribution of the market rather than only at the extremes.

The study is anchored in fiscal year 2023. This year is the final equilibrium state of the voluntary reporting era in Europe: it follows the exogenous distortion of the 2022 energy crisis, which contaminated emission and revenue data, and precedes the implementation of the Corporate Sustainability Reporting Directive (CSRD) in 2024 to 2025, which introduced mandatory, audited disclosures. The year 2023, therefore, provides the most stable baseline of strategic behaviour under voluntarism and the benchmark against which the effect of mandatory regulation can be assessed in later work.

The paper makes four contributions. First, it constructs and validates a continuous Disclosure-Performance Gap (DPG) measure for a large, multi-sector, multi-country European sample. Second, by integrating signalling theory (Spence, 1973), institutional theory (DiMaggio \& Powell, 1983; Meyer \& Rowan, 1977), and agency theory (Jensen \& Meckling, 1976), it decomposes the determinants of the gap into substantive (costly) and symbolic (costless) signal categories. Third, it extends the framework of Kim and Lyon (2015) to the European transition context, where the strategic calculus of greenwashing shifts from the avoidance of political scrutiny toward the maximisation of institutional capital attraction. Fourth, central to the paper's title, it provides a direct firm-level demonstration of proxy sensitivity: the estimated determinants of the gap are conditional on whether disclosure quality is proxied by the CDP Climate Score or the LSEG Environmental Pillar Score, holding the sample fixed.

The remainder of the paper is organised as follows. Section 2 reviews the literature, develops the integrated theoretical framework, and states the hypotheses. Section 3 describes the sample, the construction of the dependent and independent variables, and the systematic six-stage model selection process used to identify the estimated model. Section 4 reports the descriptive, multivariate, and proxy-sensitivity results. Section 5 interprets the findings, draws out their implications for investors and the post-CSRD regime, states the limitations of the design, and concludes. The appendices report the regression diagnostics and coefficient robustness, the correlation statistics, the candidate variable inventory, and the sector-specific decomposition of the estimated model.

\section*{2.\quad LITERATURE REVIEW AND THEORETICAL FRAMEWORK}

The divergence between corporate environmental disclosure and environmental performance has generated a substantial but theoretically fragmented literature. This section proceeds in four steps. It first traces the shift from binary, scandal-based measures of greenwashing to the continuous, gap-based measure adopted here. It then develops the three theoretical lenses that jointly explain why the gap arises and why it varies across firms: signalling theory as the economic lens, institutional theory as the sociological lens, and agency theory as the governance lens. It then consolidates these lenses into a single conceptual framework that maps each class of firm-level signal onto a predicted effect on the gap. It closes by stating the eight hypotheses that the empirical model tests.

\subsection*{From binary classification to continuous measurement}

The early greenwashing literature was dominated by binary or categorical measures of environmental misrepresentation. Lyon and Maxwell (2011) model greenwashing as a firm's strategic choice to selectively disclose positive environmental information under the threat of audit. Clarkson et al. (2008) provide early evidence that voluntary disclosures are positively associated with genuine performance, but only for hard, verifiable information; for discretionary narrative disclosures the relationship is ambiguous or reversed. Cho and Patten (2007) show that firms with poor environmental performance devote more disclosure space to environmental topics, an early empirical signal of talk-walk decoupling.

The conceptual anchor for the continuous approach adopted here is the typology of Delmas and Burbano (2011), who define greenwashing not as a discrete act of deception but as an intersectional state located on two continuous dimensions: the quality of environmental communication (Talk) and the quality of environmental performance (Walk). Crossing these dimensions yields four firm types. Vocal green firms combine high Talk with high Walk and represent the aligned ideal. Silent green firms combine low Talk with high Walk, a pattern sometimes termed greenhushing, in which genuine performers under-communicate. Silent brown firms combine low Talk with low Walk and correspond to traditional, openly high-emitting operators. Greenwashing firms combine high Talk with low Walk. The present study isolates the determinants that push firms toward this last quadrant. The decisive methodological consequence of adopting the intersectional definition is that position on the two axes is operationalised as a continuous distance rather than as a binary flag. A binary, litigation-based or scandal-based classification observes only the extreme tail of the distribution and therefore suffers from survivorship bias, overlooking the sub-legal, systemic decoupling that occurs across the broad middle of the market. The continuous Disclosure-Performance Gap defined in Section 3 measures exactly this distance for every firm.

The most direct predecessor to the present study is Kim and Lyon (2015), who examine U.S. corporate disclosure behaviour and identify a ``brownwashing'' phenomenon in which firms facing deregulation-minded investors understate their environmental performance to avoid political exposure. Their two-dimensional Talk-Walk matrix establishes that strategic decoupling can occur in both directions, not only inflation (greenwashing) but also deflation (brownwashing). Four divergences motivate a new empirical model. First, Kim and Lyon operate in the U.S. voluntary era, where silence was viable; the European transition context renders brownwashing strategically obsolete as the CSRD mandates comprehensive disclosure. Second, their ``Walk'' variable is the Toxic Release Inventory, a site-specific pollution measure, whereas ours is Scope 1+2 intensity, a strategic proxy for climate transition risk. Third, their ``Talk'' variable is a keyword count on corporate websites, whereas ours is the CDP Climate Score, an institutionalised framework linked to capital allocation. Fourth, their primary driver is political legitimacy, whereas ours is economic legitimacy.

The more recent literature has reinforced the case for a continuous, market-wide measure. Bingler et al. (2022) apply natural language processing to corporate climate disclosures and find pervasive ``cheap talk'', high-confidence language about ambiguous targets combined with a lack of quantitative commitments. Berg et al. (2022) supply the macro-level context: systematic ESG rating divergence creates the informational environment in which strategic decoupling is rational. If investors cannot distinguish disclosure quality from performance quality, the cost of producing high-quality disclosure is severed from its reward, creating the pooling equilibrium that Spence's (1973) framework predicts. The contribution of the present study is to carry that macro-level confusion down to the firm level and to show that it reappears inside a single greenwashing estimate.

\subsection*{Signalling theory: the economics of credible commitment}

Signalling theory, originating in Spence's (1973) analysis of labour market information asymmetry, provides the primary economic framework. The central problem is directly applicable to ESG markets: how can a high-quality firm credibly communicate its quality to an investor who cannot observe it directly? Spence's key axiom is that for a signal to be informative its cost must be negatively correlated with quality, that is, sufficiently expensive that low-quality firms find imitation prohibitive. This condition generates a separating equilibrium in which the signal reliably distinguishes high-quality from low-quality firms.

Applied to corporate environmental strategy, two categories of signal are distinguished on the cost dimension. Costly signals require substantial, irretrievable outlays with no return conditional on genuine performance improvement: retrofitting industrial processes, transitioning to renewable energy procurement, and allocating capital expenditure to emissions reduction. These commitments are physically verifiable and operationally consequential; a firm cannot produce renewable energy certificates or decommission a coal boiler through its communications team. As Connelly et al. (2011) note, signal costs must be observable and non-mimicable for the signal to function, and costly environmental signals satisfy both conditions. Costless signals, by contrast, require minimal outlays regardless of performance: endorsing a voluntary framework, publishing a net-zero pledge, or joining a sustainability index. Crawford and Sobel (1982) formalise this as ``cheap talk'', communication that is free for the sender. This condition generates a separating equilibrium in which the signal reliably distinguishes high quality from low quality firms.\footnote{This analysis motivates H1 (renewable energy) and H4 (environmental capex) as costly-signal proxies.}

\subsection*{Institutional theory: the sociology of ceremonial conformity}

While signalling theory explains disclosure through rational economic optimisation, institutional theory offers a complementary sociological account. DiMaggio and Powell's (1983) account of organisational isomorphism posits that firms facing common institutional environments converge on similar responses, not necessarily because they are efficient, but because they satisfy the normative and mimetic pressures that govern legitimate participation in a field. Three processes are relevant: coercive isomorphism (pressure from powerful stakeholders), mimetic isomorphism (imitation under uncertainty), and normative isomorphism (professionalised expectations of best practice). In ESG, all three operate simultaneously: passive investors coerce index constituents to maintain ESG scores, firms mimic disclosure leaders, and sustainability consultants professionalise the production of disclosures.

The critical implication, developed by Meyer and Rowan (1977) and elaborated by Bromley and Powell (2012), is that institutional pressure drives firms to adopt the ceremonial structures of legitimacy (policies, frameworks, targets) while decoupling those structures from operational practice. The sustainability report and the net-zero pledge become ``rationalised myths'' that confer legitimacy independent of operational content. This predicts that high-visibility firms exhibit wider gaps not despite their prominence but because of it. The role of index membership is central. Flagship index constituents are subject to disproportionate scrutiny from passive institutional investors who incorporate ESG scores into portfolio construction and proxy voting (Ioannou \& Serafeim, 2015). This creates a coercive pressure concentrated on index members to maximise disclosure scores, operating through capital allocation rather than regulatory enforcement: a firm that fails to maintain a credible disclosure score risks reallocation of passive capital.\footnote{This analysis motivates H2 (index membership) and H3 (TCFD endorsement) as costless-signal proxies.}

\subsection*{Agency theory, governance, and financial controls}

Agency theory (Jensen \& Meckling, 1976) conceptualises the manager-shareholder relationship as a principal-agent contract characterised by information asymmetry. Managers possess private information about the firm's transition risk and abatement cost structure that shareholders cannot verify. Greenwashing emerges as an agency cost when managers selectively disclose positive environmental information to secure performance-contingent compensation, protect reputational capital, or avoid the career risk of reporting poor outcomes. The theoretical remedy is monitoring through information intermediaries: analysts, auditors, and verification bodies such as the Science Based Targets initiative (SBTi). Ioannou and Serafeim (2015) document that analyst coverage is associated with improved reporting quality. However, the effectiveness of this channel depends on analysts incentivising disclosure quality (the ``Walk'') rather than disclosure quantity (the ``Talk''). If analysts reward ESG coverage rather than emission performance, they may amplify the greenwashing incentive. \footnote{This motivates H8 as Higher analyst coverage narrows the DPG} SBTi verification represents the highest available standard of voluntary target validation, but Commitment Consistency Theory (Cialdini, 1984) suggests that ambitious targets can function as legitimacy shields. The 2023 SBTi validation pause for Oil and Gas further constrains the variable in the Energy sector, as discussed in Section 3. \footnote{These limitations motivate the null test in H7 as SBTi verification narrows the DPG}.

Two further frameworks inform the governance and financial controls. Critical Mass Theory, originating in Kanter's (1977) sociology and extended by Joecks et al. (2013) and Kramer et al. (2006), posits that minority representation produces substantive change only above a critical threshold, typically three directors or 30 per cent of board composition; below it, minority members function as tokens. Applied to environmental governance, H5 (Higher board gender diversity narrows the DPG) tests whether board gender diversity produces genuine alignment or symbolic diversity-washing. Slack Resources Theory (Bourgeois, 1981; Waddock \& Graves, 1997) offers a capability argument for H6 (Higher profitability (ROA) narrows the DPG): profitable firms possess discretionary resources for costly decarbonisation, so under this theory greenwashing should be inversely related to profitability. Rejection of H6 would imply that greenwashing is a strategic choice available to and exercised by profitable firms rather than a symptom of financial distress.

\subsection*{A unified conceptual framework}

The three lenses are complementary rather than competing, and they consolidate into a single framework that organises the firm-level determinants of the gap by the cost structure of the signal involved. Figure~\ref{fig:framework} sets out this framework. It models the alignment between Talk and Walk as a rational economic outcome rather than as a moral attribute of the firm, and it distinguishes three pathways. Along the substantive pathway, costly signals, namely renewable energy use and environmental capital expenditure, require irretrievable financial outlays and therefore act as credible proxies for genuine performance, producing a separating equilibrium in which the Walk matches the Talk and the gap narrows. Along the symbolic pathway, low-cost visibility signals, namely flagship index membership and TCFD support, are driven by coercive isomorphism and legitimacy pressure; because they are cheap to adopt but highly visible, they incentivise firms to maximise disclosure without altering operations, and the gap widens. Along the moderating pathway, internal governance (board diversity, financial slack) and external verification (SBTi validation, analyst coverage) are hypothesised to act as brakes on decoupling by raising the efficacy of monitoring; their failure to narrow the gap would indicate a breakdown of the voluntary governance ecosystem. The framework therefore predicts that the gap widens precisely when the market rewards symbolic signals more than substantive ones.

\begin{figure}[H]
\caption{Unified conceptual framework}
\label{fig:framework}
\centering
\resizebox{\textwidth}{!}{%
\begin{tikzpicture}[
  inbox/.style={rectangle, rounded corners, draw=black!70, fill=black!4, text width=3.5cm, align=center, minimum height=1.7cm, font=\footnotesize},
  mech/.style={rectangle, rounded corners, draw=black!70, fill=white, text width=3.4cm, align=center, minimum height=1.5cm, font=\footnotesize},
  eff/.style={rectangle, rounded corners, draw=black!70, text width=3.0cm, align=center, minimum height=1.3cm, font=\footnotesize},
  dpg/.style={rectangle, rounded corners, draw=black, very thick, fill=black!8, text width=3.2cm, align=center, minimum height=6.2cm, font=\small},
  arr/.style={-{Latex[length=2.4mm]}, thick, draw=black!70}
]
\node[inbox] (s1) at (0,3.4) {\textbf{Substantive (costly) signals}\\ Renewable energy\\ Environmental capex};
\node[inbox] (s2) at (0,0) {\textbf{Symbolic (visibility) signals}\\ Index membership\\ TCFD support};
\node[inbox] (s3) at (0,-3.4) {\textbf{Moderating controls}\\ Board diversity, financial slack\\ SBTi, analyst coverage};
\node[mech] (m1) at (5,3.4) {\textbf{Signalling theory}\\ Costly, irretrievable outlay $\rightarrow$ separating equilibrium};
\node[mech] (m2) at (5,0) {\textbf{Institutional theory}\\ Coercive isomorphism; policy-practice decoupling};
\node[mech] (m3) at (5,-3.4) {\textbf{Agency \& critical mass theory}\\ Monitoring as a brake};
\node[eff, fill=black!4] (e1) at (10,3.4) {Narrows the gap\\ ($-$)};
\node[eff, fill=black!10] (e2) at (10,0) {Widens the gap\\ ($+$, greenwashing)};
\node[eff, fill=white] (e3) at (10,-3.4) {Hypothesised brake;\\ found ineffective};
\node[dpg] (d) at (14,0) {\textbf{Disclosure-\\ Performance Gap}\\[4pt] $Z(\text{Talk}) - Z(\text{Walk})$\\[4pt] \footnotesize Talk: CDP score\\ Walk: inverted Scope 1+2 intensity};
\draw[arr] (s1) -- (m1); \draw[arr] (m1) -- (e1); \draw[arr] (e1) -- (d);
\draw[arr] (s2) -- (m2); \draw[arr] (m2) -- (e2); \draw[arr] (e2) -- (d);
\draw[arr] (s3) -- (m3); \draw[arr] (m3) -- (e3); \draw[arr] (e3) -- (d);
\end{tikzpicture}%
}
\footnotesize\textit{Source: Author's own elaboration, integrating Spence (1973), DiMaggio and Powell (1983), and Jensen and Meckling (1976)}
\end{figure}

The alignment between Talk (CDP disclosure) and Walk (emission performance) is modelled as a function of the cost structure of the signals a firm sends. Costly substantive signals narrow the gap through a separating equilibrium; low-cost symbolic signals widen it through ceremonial conformity; governance and verification controls are hypothesised to act as brakes. Reconstructed from the unified framework developed in the underlying thesis.

To test the aforementioned theories, we define eight hypotheses that are summarised in Table 1.

\begin{table}[H]
\centering
\textbf{Table 1. Hypothesis summary by theoretical dimension}\\[4pt]
\begin{tabular}{@{}
  >{\raggedright\arraybackslash}p{(\columnwidth - 6\tabcolsep) * \real{0.06}}
  >{\raggedright\arraybackslash}p{(\columnwidth - 6\tabcolsep) * \real{0.40}}
  >{\raggedright\arraybackslash}p{(\columnwidth - 6\tabcolsep) * \real{0.30}}
  >{\raggedright\arraybackslash}p{(\columnwidth - 6\tabcolsep) * \real{0.24}}@{}}
\toprule
\textbf{H} & \textbf{Hypothesis statement} & \textbf{Theoretical basis} & \textbf{Predicted / Result} \\
\midrule

H1 & Higher renewable energy use narrows the DPG & Signalling: costly signal & Negative / Supported*** \\
H2 & Flagship index membership widens the DPG & Institutional: coercive isomorphism & Positive / Supported*** \\
H3 & Voluntary TCFD endorsement widens the DPG & Institutional: ceremonial conformity & Positive / Supported** \\
H4 & Higher environmental capex narrows the DPG & Signalling: costly signal & Negative / Supported** \\
H5 & Higher board gender diversity narrows the DPG & Critical Mass Theory & Negative / Not selected \\
H6 & Higher profitability (ROA) narrows the DPG & Slack Resources Theory & Negative / Not selected \\
H7 & SBTi verification narrows the DPG & Agency: external verification & Negative / Not selected \\
H8 & Higher analyst coverage narrows the DPG & Agency: information intermediaries & Negative / Not selected \\
\bottomrule
\end{tabular}

\noindent\begin{minipage}{\linewidth}\footnotesize Note: DPG = Disclosure-Performance Gap. A positive DPG indicates greenwashing (Talk \textgreater{} Walk); a negative DPG indicates brownwashing. Significance levels are those of the estimated primary model. *** p \textless{} 0.01, ** p \textless{} 0.05, * p \textless{} 0.10. Source: Author's own elaboration\end{minipage}\normalsize
\end{table}

\section*{3.\quad DATA AND METHODOLOGY}

This section sets out the empirical design in three steps. It first describes the sample and the rationale for restricting the analysis to four carbon-intensive sectors in a single fiscal year. It then defines the dependent variable and the candidate independent variables, including the standardisation that places disclosure and performance on a common scale. It finally describes the six-stage process that reduces a large candidate set to the estimated model, together with the estimator used for inference. The full inventory of candidate variables is reported in Appendix C.

\subsection*{Research design and sample construction}

The study uses a cross-sectional design moving from established theory to testable hypotheses rather than building theory from the data. A single cross-section is the right choice here because the research question is about where firms sit in equilibrium within the voluntary disclosure era, not about how they change over time. The underlying variables, carbon intensity, capital expenditure, and governance composition, are observable from public records and exist independently of any modelling choice.

The results are associational, not casual, and should be read as such. One cross-section cannot rule out reverse causality: firms that already disclose in line with their performance may simply be more inclined to invest in renewables, rather than the investing driving the disclosure gap. Nor can it rule out unobserved factors, such as how long a firm has operated in low-carbon markets or how easily it can access transition capital, that could influence both the signals and the gap simultaneously. The coefficients therefore measure conditional associations, holding the other variables fixed, not treatment effects. 

The sample is constructed based on STOXX Europe 600 Index, which covers large-, mid-, and small-capitalisation firms across 17 European countries and is the primary universe from which European institutional capital is allocated. Its constituents attract the highest levels of ESG-linked scrutiny, making it the most relevant population for testing the institutional isomorphism hypotheses.

To maximise internal validity, the sample is restricted to four high-impact sectors: Energy (GICS 10), Materials (GICS 15), Industrials (GICS 20), and Utilities (GICS 55). Service-oriented sectors are excluded because their direct (Scope 1 and 2) emissions are structurally negligible, which makes the ``Walk'' variable uninterpretable as a proxy for decarbonisation. The four retained sectors account for the large majority of industrial greenhouse gas emissions and represent the arena where the tension between costly decarbonisation and symbolic disclosure is most acute. To prevent sector-dominance bias, a balanced sampling design selects the top 50 firms by market capitalisation within each sector, yielding an initial panel of 200 firms. From 219 in-scope candidates, the retained firms are those carrying both a 2023 CDP score and an LSEG environmental pillar score; firms receiving a CDP ``F'' (failure to disclose) are retained and coded zero, preserving the strategic signal of deliberate non-disclosure rather than treating silence as missing data.

The observation year is 2023, chosen to isolate strategic behaviour under voluntary reporting at its most analytically clean. The years 2020 to 2022 data are contaminated by pandemic-induced emission reductions and by the 2022 energy crisis, which distorted energy consumption, commodity-linked revenue, and capital allocation. The years 2024 to 2025 data exhibit transitional instability as firms adapt to CSRD requirements, introducing structural breaks in both Talk and Walk. Year 2023 occupies the unique position of a strategic equilibrium year, the final state of voluntary behaviour before mandatory obligations alter the incentive structure.

To minimise measurement error and common-method bias, data were not drawn from a single aggregator. Every data point was triangulated under a strict source hierarchy that prioritises firm-reported primary data, summarised in Table 2.

\begin{table}[!ht]
\centering
\textbf{Table 2. Data source hierarchy for variable triangulation}\\[4pt]
\begin{tabular}{@{}
  >{\raggedright\arraybackslash}p{(\columnwidth - 4\tabcolsep) * \real{0.16}}
  >{\raggedright\arraybackslash}p{(\columnwidth - 4\tabcolsep) * \real{0.52}}
  >{\raggedright\arraybackslash}p{(\columnwidth - 4\tabcolsep) * \real{0.32}}@{}}
\toprule
\textbf{Priority} & \textbf{Source type and examples} & \textbf{Role} \\
\midrule

Primary & Company official filings: annual, integrated, sustainability, and corporate governance reports, and non-financial information statements (Year 2023) & Default source for financial, governance, and operational environmental data reported directly by the firm. \\
Primary & Established ESG databases: CDP Climate Change Programme (scores and emissions), SBTi dashboard, LSEG (Refinitiv) environmental pillar, FTSE4Good, DJSI constituent lists & Default source for ESG-specific variables assessed under standardised methodologies. \\
Primary & Tracker databases: archived TCFD supporters list, Net Zero Tracker & Default source for binary commitment variables. \\
Secondary & Financial aggregators and regulatory filings: market data vendors, ECB reference rates, local securities regulators and exchanges & Market data, currency conversion, analyst coverage, and verification of firm-reported figures. \\
Tertiary & Documented proxy methods: EU taxonomy-eligible capex as a proxy for environmental capex; prior-year values with adjustment & Used only when primary and secondary sources are unavailable; each proxy documented in the audit trail. \\
\bottomrule
\end{tabular}
\footnotesize\textit{Source: Author's own elaboration}
\end{table}

\subsection*{Variable construction}

The choice of disclosure proxy is treated as an explicit analytical decision rather than an exclusion. The CDP Climate Score is adopted as the primary measure of Talk because it captures the firm's own questionnaire response to an institutionalised framework rather than a rating agency's algorithmic transformation of that response. The LSEG Environmental Pillar Score, a third-party assessment, is retained as an alternative disclosure proxy and is used in Section 4 to re-estimate the model on the identical sample. This design converts the proxy question into a test: if the determinants of the gap are stable, they should survive the substitution of one institutional proxy for another; if they are not, that instability is itself evidence of aggregate confusion.

\clearpage
\textit{Dependent variable:} The central modelling problem is that disclosure quality (CDP Score) is ordinal while emission performance is a continuous ratio. Direct comparison is invalid, so both components are placed on commensurate units by within-sector Z-standardisation. The CDP Climate Score for the year 2023 is converted to a numeric scale (A=8, A-=7, B=6, B-=5, C=4, C-=3, D=2, D-=1, F=0), with non-respondents coded zero, then standardised within sector:
\begin{equation}
Z(\mathrm{Talk})_{i} = \frac{\mathrm{CDP}_{i} - \mu^{(s)}_{\mathrm{CDP}}}{\sigma^{(s)}_{\mathrm{CDP}}}
\end{equation}
Emission performance is Scope 1 + Scope 2 intensity, total tonnes of CO\textsubscript{2} equivalent divided by Year 2023 revenue. Revenue normalisation controls for scale. The sign is inverted so that a higher score reflects better performance:
\begin{equation}
Z(\mathrm{Walk})_{i} = -\,\frac{\mathrm{Emission}_{i} - \mu^{(s)}_{\mathrm{Emission}}}{\sigma^{(s)}_{\mathrm{Emission}}}
\end{equation}
The gap is the difference of the two standardised components:
\begin{equation}
\mathrm{DPG}_{i} = Z(\mathrm{Talk})_{i} - Z(\mathrm{Walk})_{i}
\end{equation}
A positive DPG indicates that disclosure exceeds performance (greenwashing); a negative DPG indicates the reverse (brownwashing); a value near zero indicates alignment. Because both components are standardised within sector, each sector mean of the DPG is zero by construction, a property that governs the interpretation of the sector dummies below.

The Talk component requires one measurement assumption that is stated explicitly. The CDP grade is ordinal, and assigning equally spaced integers (A=8 down to F=0) before standardising treats successive grades as equal intervals. This is a common and defensible simplification, because the within-sector standardisation rescales the coded grade by the sector dispersion rather than relying on the raw spacing, and because the binary distinction that carries most of the signal, disclosure versus non-disclosure, is preserved regardless of spacing. The assumption is addressed directly in the robustness analysis of Section 4: a rank-based reconstruction of the gap, which replaces the cardinal coding with within-sector ranks, leaves every coefficient unchanged in sign and significance, so the equal-interval simplification does not affect the conclusions.

\textit{Candidate independent variables:} The four variables selected into the final model are defined as follows. The two continuous economic signals are standardised across the sample to permit comparison of effect magnitudes; for a continuous variable $X$,
\begin{equation}
X^{\mathrm{Std}}_{i} = \frac{X_{i} - \mu_{X}}{\sigma_{X}} .
\end{equation}
Renewable energy use is the percentage of total energy consumption from renewable sources in year 2023, $\mathrm{Renew}_{i}$, entered as $\mathrm{Renew}^{\mathrm{Std}}_{i} = (\mathrm{Renew}_{i}-\mu_{\mathrm{Renew}})/\sigma_{\mathrm{Renew}}$ (H1, operational OPEX commitment). Environmental capital expenditure is the share of total capex allocated to taxonomy-aligned environmental projects, $\mathrm{Capex}_{i}$, entered as $\mathrm{Capex}^{\mathrm{Std}}_{i} = (\mathrm{Capex}_{i}-\mu_{\mathrm{Capex}})/\sigma_{\mathrm{Capex}}$ (H4, forward-looking CAPEX commitment). The two symbolic signals are retained in raw binary form to preserve interpretation: index membership $\mathrm{Index}_{i}\in\{0,1\}$ equals one if the firm is a constituent of the FTSE 100, DAX 40, CAC 40, or IBEX 35 in Year 2023 (H2), and TCFD support $\mathrm{TCFD}_{i}\in\{0,1\}$ equals one if the firm was registered on the archived TCFD supporters database at its dissolution in November 2023 (H3). Sector membership enters as three dummies (Industrials, Materials, Utilities) with Energy as the reference category.

Governance and financial controls tested but not selected are board gender diversity (percentage of board seats held by women, H5), return on assets (standardised, H6), SBTi validation status (binary, H7; structurally zero for Energy owing to the 2023 sector pause), and analyst coverage (number of covering analysts, H8). Firm size (log total assets) was tested as a control. Revenue growth and capital intensity were excluded after screening: revenue growth carried severe commodity-inflation noise in Energy and Utilities, and capital intensity introduced collinearity given the sector restriction. Table 3 summarises the definition and source of every variable entering the analysis.

\begin{table}[H]
\centering
\textbf{Table 3. Definition and source of model variables}\\[4pt]
\begin{tabular}{@{}
  >{\raggedright\arraybackslash}p{(\columnwidth - 4\tabcolsep) * \real{0.30}}
  >{\raggedright\arraybackslash}p{(\columnwidth - 4\tabcolsep) * \real{0.70}}@{}}
\toprule
\textbf{Variable} & \textbf{Definition and source} \\
\midrule

DPG (dependent) & $Z(\text{Talk}) - Z(\text{Walk})$, both standardised within sector. Talk = CDP Climate Score (A=8 to F=0); Walk = inverted Scope 1+2 intensity per revenue, Year 2023. \\
\% Renewable Energy (Std) & Renewable share of total energy consumption, standardised. Source: sustainability reports, CDP responses. \\
Index Membership (0/1) & 1 if Year 2023 constituent of FTSE 100, DAX 40, CAC 40, or IBEX 35. Source: official index lists. \\
TCFD Supporter (0/1) & 1 if registered on the archived TCFD supporters database at dissolution, November 2023. \\
\% Environmental Capex (Std) & Taxonomy-aligned environmental capex as a share of total capex, standardised. Source: annual reports, EU taxonomy appendices. \\
Sector dummies & Industrials, Materials, Utilities; Energy is the reference category. \\
Alternative Talk (robustness) & LSEG (Refinitiv) Environmental Pillar Score, used in the proxy-substitution test in Section 4. \\
\bottomrule
\end{tabular}
\footnotesize\textit{Source: Author's own elaboration}
\end{table}

\subsection*{Model selection and estimation}

\begin{figure}[H]
\centering
\caption{Stepwise AICc progression across the forward-selection search.}
\label{fig:stepwise}
\includegraphics[width=0.8\textwidth]{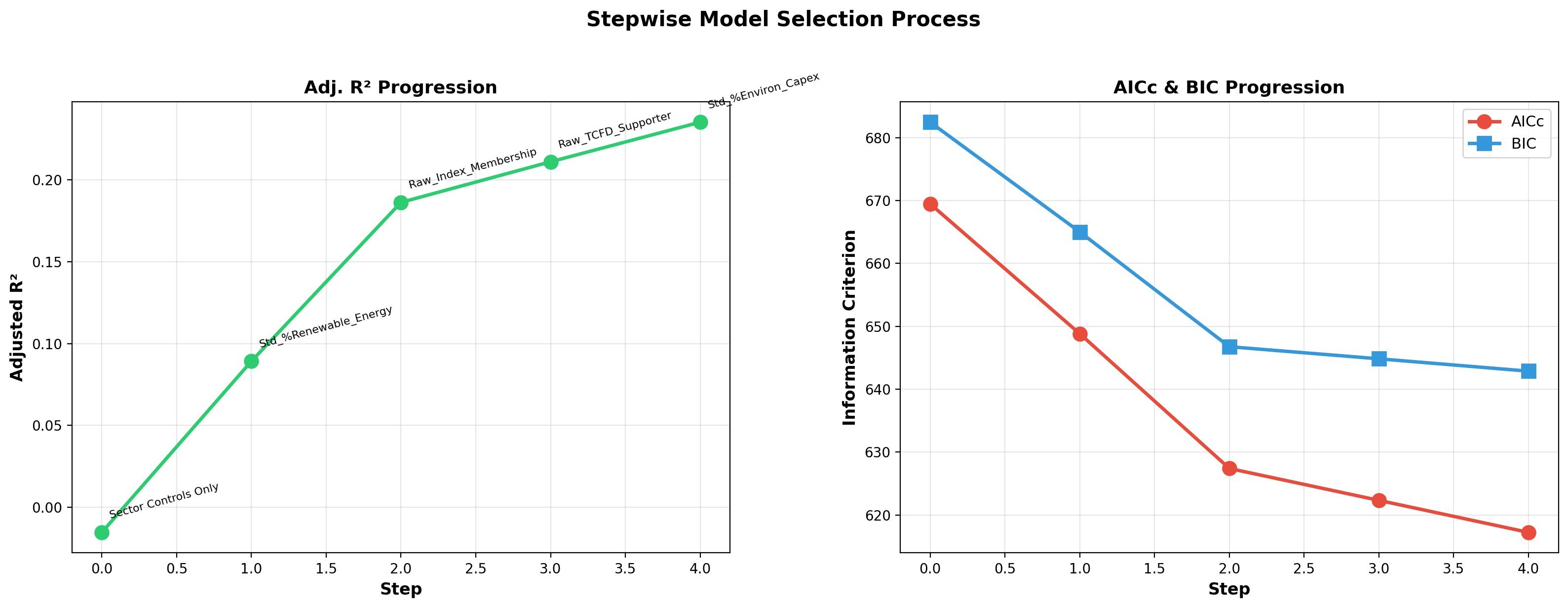}
\begin{center}
    \footnotesize\textit{Source: Author's own elaboration}
\end{center}
\end{figure}

\begin{figure}[H]
\caption{Variance Inflation Factors across candidate pools; all values are well below the threshold of 5.0.}
\label{fig:vifpools}
\centering
\includegraphics[width=0.8\textwidth]{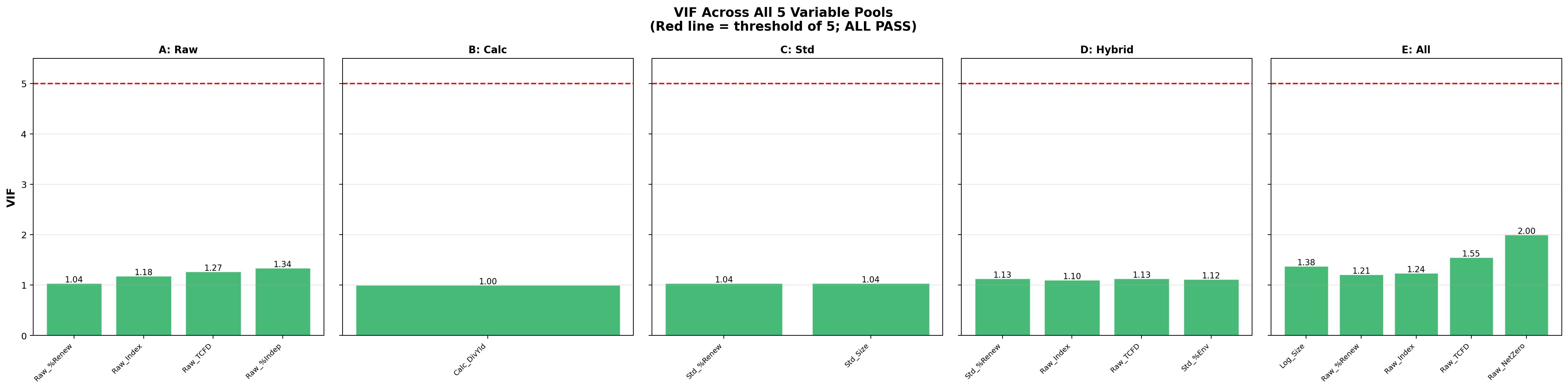}
\begin{center}
    \footnotesize\textit{Source: Author's own elaboration}
\end{center}
\end{figure}

Variable selection follows a systematic six-stage process designed to limit researcher discretion and the risk of p-hacking. The protocol and its outputs are set out in full in this section. Stage 1 assembles 36 candidate variables spanning financial, operational, governance, and disclosure metrics; the full inventory, grouped by type, is listed in Appendix C. Stage 2 ranks them by Pearson correlation with the DPG to identify preliminary predictors. Stage 3 removes any variable with a Variance Inflation Factor (VIF) above 5.0. Stage 4 applies a stepwise forward search across five distinct variable pools (raw, calculated, standardised, hybrid, and all-variables) with the corrected Akaike Information Criterion (AICc) as the objective; this stepwise search spans 421 candidate specifications and is the core of the procedure. Because the AICc penalises complexity, hypothesised predictors that fail to improve information quality, including board diversity, return on assets, and analyst coverage, are pruned automatically rather than by judgement. Stage 5 screens the winning specification for influential observations using Cook's distance. Stage 6 re-estimates the final model with HC3 robust standard errors to correct for heteroscedasticity. Table 4 records the funnel from the full candidate set to the estimated model. Figure~\ref{fig:stepwise} traces the AICc path of the forward search, and Figure~\ref{fig:vifpools} confirms that multicollinearity is controlled across all candidate pools.

\begin{table}[!ht]
\centering
\textbf{Table 4. Model selection funnel, by stage}\\[4pt]
\begin{tabular}{@{}
  >{\raggedright\arraybackslash}p{(\columnwidth - 8\tabcolsep) * \real{0.06}}
  >{\raggedright\arraybackslash}p{(\columnwidth - 8\tabcolsep) * \real{0.26}}
  >{\raggedright\arraybackslash}p{(\columnwidth - 8\tabcolsep) * \real{0.40}}
  >{\raggedright\arraybackslash}p{(\columnwidth - 8\tabcolsep) * \real{0.28}}@{}}
\toprule
\textbf{Stage} & \textbf{Operation} & \textbf{Criterion} & \textbf{Carried forward} \\
\midrule

1 & Candidate assembly & Theory and data availability & 36 candidate variables \\
2 & Bivariate screening & Pearson correlation with the DPG; 20 single-variable regressions & 20 ranked predictors \\
3 & Multicollinearity filter & VIF \textgreater{} 5.0 and $|r| \textgreater 0.70$ pairs removed; duplicate transforms dropped & Non-redundant pool of 14 to 23 variables per scale type \\
4 & Stepwise pool search & Forward and bidirectional stepwise under AICc, across five variable pools (421 specifications) & 4 predictors plus 3 sector dummies (Pool D) \\
5 & Influence screening & Cook's distance $D > 4/n$ (n=200) & 16 firms flagged; n=200 retained as primary, n=184 as robustness \\
6 & Robust estimation & HC3 heteroscedasticity-consistent standard errors & Final reported model \\
\bottomrule
\end{tabular}

\noindent\begin{minipage}{\linewidth}\footnotesize Note: stages 1 to 4 operate on the variable space; stage 5 operates on observations. The 421 specifications in stage 4 are the total stepwise paths evaluated across the five pools. Source: Author's own elaboration \end{minipage}\normalsize
\end{table}

\begin{figure}[H]
\caption{Information-criterion comparison across the five variable pools.}
\label{fig:poolcomp}
\centering
\includegraphics[width=0.8\textwidth]{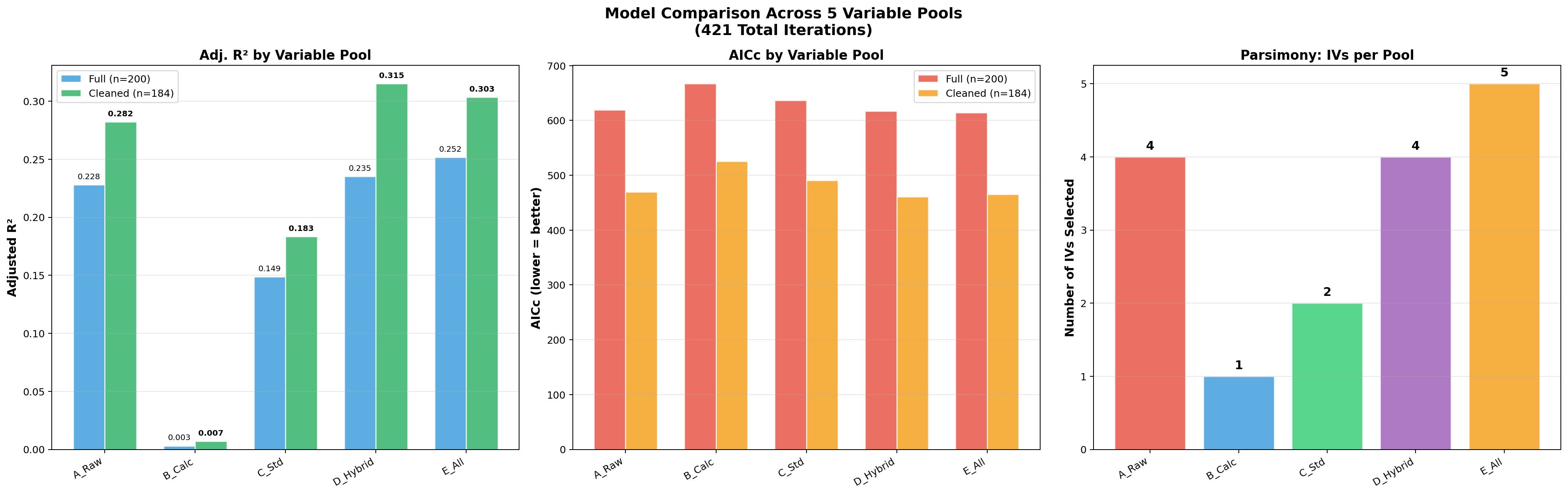}
\begin{center}
    \footnotesize\textit{Source: Author's own elaboration}
\end{center}
\end{figure}

\clearpage
The pool comparison does not point to a single dominant model on information criteria alone. On the full sample the all-variables pool (Pool E) attains the lowest AICc (614.01), marginally below the hybrid pool (Pool D, 617.24) and the raw pool (Pool A, 619.12). Pool E, however, mixes standardised and raw scales, which impairs coefficient interpretation, and it does not retain its advantage on the influence-trimmed sample. The hybrid specification (Pool D), which combines standardised continuous economic variables with raw binary governance variables, is selected on two grounds: it is the most interpretable, because standardised coefficients read as one-standard-deviation effects while binaries read as zero-to-one effects, and it attains the best adjusted $R^{2}$ on the cleaned sample (0.315). AICc is used to compare pools only within the same sample, never across the full and trimmed samples. The selected model is therefore chosen for interpretability and stability rather than for a marginal AICc advantage on the full sample. Table 5 reports the full five-pool comparison, and Figure~\ref{fig:poolcomp} displays the same information graphically.

\begin{table}[!ht]
\centering
\textbf{Table 5. Five-pool model comparison (full sample, n=200)}\\[4pt]
\begin{tabular}{@{}
  >{\raggedright\arraybackslash}p{(\columnwidth - 8\tabcolsep) * \real{0.18}}
  >{\raggedright\arraybackslash}p{(\columnwidth - 8\tabcolsep) * \real{0.26}}
  >{\raggedright\arraybackslash}p{(\columnwidth - 8\tabcolsep) * \real{0.10}}
  >{\raggedright\arraybackslash}p{(\columnwidth - 8\tabcolsep) * \real{0.13}}
  >{\raggedright\arraybackslash}p{(\columnwidth - 8\tabcolsep) * \real{0.15}}
  >{\raggedright\arraybackslash}p{(\columnwidth - 8\tabcolsep) * \real{0.13}}@{}}
\toprule
\textbf{Pool} & \textbf{Variable types} & \textbf{IVs} & \textbf{$R^2$} & \textbf{Adj.\ $R^2$} & \textbf{AICc} \\
\midrule

A (Raw) & Raw only & 4 & 0.255 & 0.228 & 619.12 \\
B (Calc) & Calculated only & 1 & 0.023 & 0.003 & 666.89 \\
C (Std) & Standardised only & 2 & 0.170 & 0.149 & 636.39 \\
\textbf{D (Hybrid, chosen)} & Std + raw binary & 4 & 0.262 & 0.235 & 617.24 \\
E (All) & All types & 5 & 0.282 & 0.252 & 614.01 \\
\bottomrule
\end{tabular}

\noindent\begin{minipage}{\linewidth}\footnotesize Note: AICc compares pools within the full sample only. Pool D is selected for interpretability and its best adjusted $R^{2}$ on the cleaned sample (0.315), not for a full-sample AICc advantage. Source: Author's own elaboration\end{minipage}\normalsize
\end{table}

The estimated specification, with Energy as the reference sector, is
\begin{equation}
\mathrm{DPG}_{i} = \beta_{0} + \beta_{1}\,\mathrm{Renew}^{\mathrm{Std}}_{i} + \beta_{2}\,\mathrm{Index}_{i} + \beta_{3}\,\mathrm{TCFD}_{i} + \beta_{4}\,\mathrm{Capex}^{\mathrm{Std}}_{i} + \sum_{s}\gamma_{s}\,\mathrm{Sector}_{s,i} + \varepsilon_{i}.
\end{equation}
The stepwise algorithm excluded board diversity, ROA, SBTi validation, and analyst coverage. This exclusion is itself a substantive result: the complexity penalty removes these variables because they add no explanatory power once substantive and symbolic signals are present, not because they were omitted by oversight. A potential concern with any stepwise procedure is that the retained set is an artefact of data mining. That concern is mitigated here because the four selected predictors, renewable energy, environmental capex, index membership, and TCFD support, are exactly the four signals nominated ex ante by the signalling and institutional hypotheses (H1 to H4). The selection therefore confirms a theory-specified model rather than discovering an atheoretical one; the data-driven step removes the governance and financial controls (H5 to H8) that theory treated as secondary. The sector-specific decomposition of equation (5) is reported in Appendix D.

\textit{Outlier treatment and the primary sample.} The full sample of 200 firms was screened with Cook's distance at the conventional threshold $D > 4/n$ ($n=200$, threshold 0.020), identifying 16 influential observations with extreme emission intensities, anomalous financial ratios, or extreme CDP scores relative to sector peers. Removing them resolves the normality failure of the full-sample residuals (Jarque-Bera falls from 49.66, p \textless{} 0.001, to 1.76, p = 0.414). The full sample of 200 firms is retained as the primary specification and is estimated with HC3 robust standard errors, which remain valid under non-normality at this sample size. The influence-trimmed sample of 184 firms is carried forward as a robustness check, not as the main model.

\section*{4.\quad RESULTS}

This section reports the empirical findings in the order in which a reader needs them. It first profiles the sample and the distribution of the dependent variable, then establishes the bivariate associations that anticipate the multivariate results, verifies the ordinary least squares assumptions, and reports the primary hybrid regression with its influence-trimmed robustness counterpart. It closes with the proxy-sensitivity test, in which the disclosure measure is replaced on the identical firms. Interpretation of mechanisms and the policy implications are reserved for Section 5, so that the evidence is presented here without editorial overlay.

\subsection*{Descriptive statistics and bivariate associations}

\begin{figure}[H]
\centering
\caption{Distribution of the Disclosure-Performance Gap by sector}\label{fig:dvdist}
\includegraphics[width=0.7\textwidth]{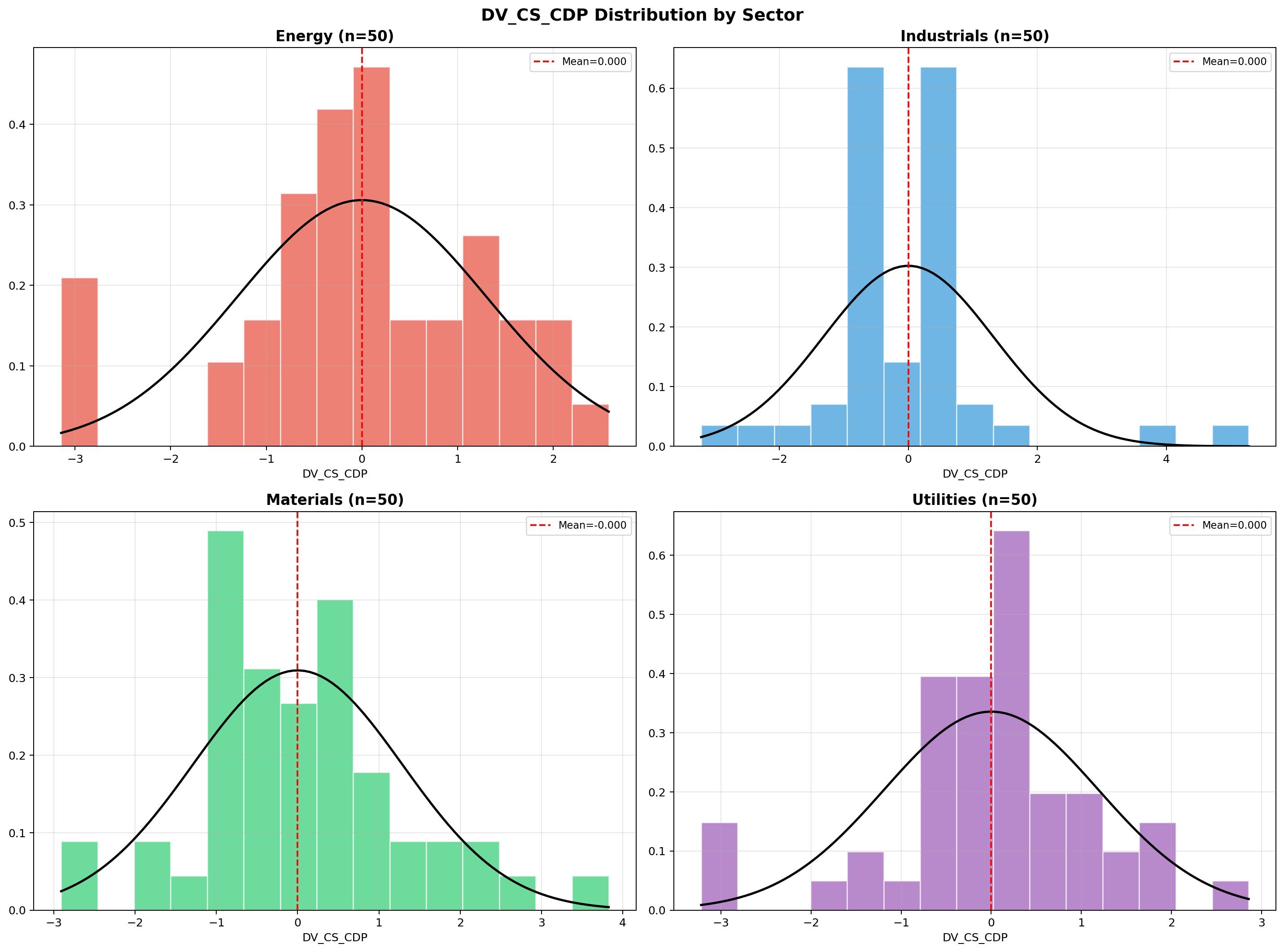}
\noindent\begin{minipage}{\linewidth}\footnotesize Note: Within-sector standardisation pins each sector mean at zero; the panels display dispersion and tail behaviour, not cross-sector greenwashing rankings. Source: Author's own elaboration\end{minipage}\normalsize
\end{figure}

The aggregate DPG has a mean of zero and a standard deviation of 1.267 by construction, ranging from $-3.22$ (extreme brownwashing) to $+5.28$ (extreme greenwashing). The distribution is mildly right-skewed (0.21) and leptokurtic (excess kurtosis 2.34): most firms cluster near zero, but the tails are heavier than under a normal distribution, and the right tail of sophisticated greenwashers extends further than the left tail of silent performers. Sector medians (Energy +0.031, Industrials +0.016, Materials $-0.029$, Utilities +0.097) are mechanical artefacts of within-sector demeaning and carry no substantive interpretation; the only distributional feature of interest is dispersion, which is the widest in Energy (SD 1.305), reflecting the split between legacy operators and transitioning producers. Figure~\ref{fig:dvdist} shows the sector-level distribution. Table 6 reports the descriptive statistics.

\begin{table}[!ht]
\centering
\textbf{Table 6. Descriptive statistics of the dependent variable and key independent variables}\\[4pt]
\begin{tabular}{@{}
  >{\raggedright\arraybackslash}p{(\columnwidth - 12\tabcolsep) * \real{0.24}}
  >{\raggedright\arraybackslash}p{(\columnwidth - 12\tabcolsep) * \real{0.12}}
  >{\raggedright\arraybackslash}p{(\columnwidth - 12\tabcolsep) * \real{0.12}}
  >{\raggedright\arraybackslash}p{(\columnwidth - 12\tabcolsep) * \real{0.12}}
  >{\raggedright\arraybackslash}p{(\columnwidth - 12\tabcolsep) * \real{0.12}}
  >{\raggedright\arraybackslash}p{(\columnwidth - 12\tabcolsep) * \real{0.12}}
  >{\raggedright\arraybackslash}p{(\columnwidth - 12\tabcolsep) * \real{0.13}}@{}}
\toprule
\textbf{Variable} & \textbf{Mean} & \textbf{SD} & \textbf{Min} & \textbf{Median} & \textbf{Max} & \textbf{Skew / Kurt} \\
\midrule

DPG (full sample) & 0.000 & 1.267 & $-3.219$ & 0.024 & 5.278 & 0.21 / 2.34 \\
DPG, Energy & 0.000 & 1.305 & $-3.145$ & +0.031 & 2.582 & --- \\
DPG, Industrials & 0.000 & 1.319 & $-3.213$ & +0.016 & 5.278 & --- \\
DPG, Materials & 0.000 & 1.290 & $-2.910$ & $-0.029$ & 3.828 & --- \\
DPG, Utilities & 0.000 & 1.189 & $-3.219$ & +0.097 & 2.858 & --- \\
Renewable energy (\%) & 30.8 & 28.4 & 0.0 & --- & 100 & --- \\
Index membership (\% of sample) & $\approx$55 & --- & --- & --- & --- & --- \\
TCFD supporter (\% of sample) & $\approx$91 & --- & --- & --- & --- & --- \\
\bottomrule
\end{tabular}

\noindent\begin{minipage}{\linewidth}\footnotesize Note: DPG is constructed as $Z(\text{CDP Score}) - Z(\text{inverted Scope 1+2 intensity})$, standardised within sector. Index and TCFD figures are approximate sample proportions. Sector-level DPG statistics are mean-zero by construction. Source: Author's own elaboration\end{minipage}\normalsize
\end{table}

The independent variables reveal a sharp split between substantive and symbolic signals. Renewable energy use averages 30.8 per cent with a high standard deviation of 28.4 per cent, indicating a polarised market with leaders above 80 per cent and a long tail of laggards below 5 per cent. Environmental capex shows similarly high variance. The symbolic signals, by contrast, are highly saturated: roughly 55 per cent of the sample are flagship index constituents and roughly 91 per cent are registered TCFD supporters, leaving only 19 non-supporting firms. The thinness of this base is relevant to the interpretation of the TCFD coefficient below.

\begin{figure}[H]
\centering
\caption{Disclosure-Performance Gap by sector}\label{fig:boxplot}
\includegraphics[width=0.6\textwidth]{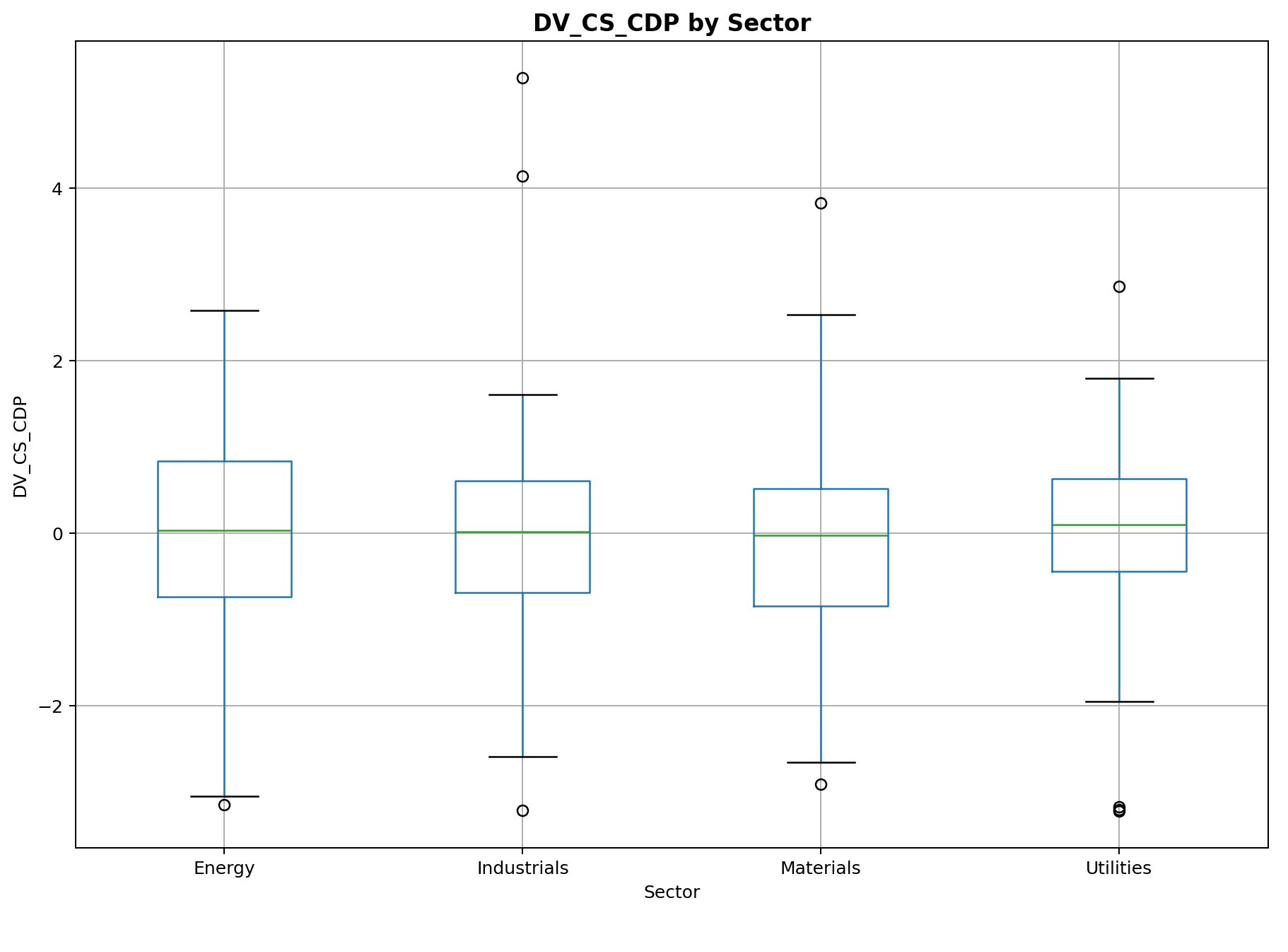}
\noindent\begin{minipage}{\linewidth}\footnotesize Note: Medians are pinned near zero by within-sector standardisation; the boxes and whiskers display dispersion and tail behaviour, not a cross-sector greenwashing ranking. Source: Author's own elaboration\end{minipage}\normalsize
\end{figure}

Figure~\ref{fig:boxplot} presents the dependent variable as boxplots by sector. All four sector medians sit at or near zero, a direct consequence of within-sector standardisation, while the interquartile range and the upper outliers are widest in Energy and Industrials, which locates the most extreme greenwashing cases in those two sectors.

Bivariate Pearson correlations track the theoretical predictions. Among substantive signals, renewable energy has the strongest negative correlation with the DPG (r = $-0.33$, p \textless{} 0.01) and environmental capex a more modest one (r = $-0.19$, p \textless{} 0.01). Among symbolic signals, index membership has the strongest positive correlation (r = +0.31, p \textless{} 0.01), followed by TCFD endorsement (r = +0.28) and analyst coverage (r = +0.20). The positive analyst correlation is a preliminary signal of monitoring herding rather than monitoring effectiveness. Governance and financial controls (board diversity, ROA, SBTi) all show correlations near zero ($|r| < 0.10$). The highest cross-predictor correlation in the final specification is between firm size and analyst coverage (r = 0.60), below the $|r| > 0.70$ exclusion threshold, and the maximum VIF across candidates is 1.18, confirming the absence of multicollinearity. The correlation matrix is reported in Appendix B.

\subsection*{Determinants of the Disclosure-Performance Gap}

Table 7 reports the final hybrid model, estimated by ordinary least squares with HC3 robust standard errors on the full sample of 200 firms; the assumption tests that justify this estimator are reported in the following subsection. The influence-trimmed sample of 184 firms is shown alongside as a robustness column. Every structural predictor is materially identical in sign, magnitude, and significance across the two samples, which establishes that the findings are not artefacts of the influential observations. The primary model explains 26.2 per cent of the variance in the gap (adjusted $R^{2}$ = 0.235), with joint significance F(7,192) = 9.74, p \textless{} 0.001. Coefficients are read on two scales by design of the hybrid specification: for the two standardised economic variables, a coefficient is the change in the gap, in standard deviations, associated with a one-standard-deviation increase in the predictor; for the two binary signals and the sector dummies, a coefficient is the gap difference between the two states, in standard deviations, holding the other predictors fixed.

The results separate cleanly into substantive and symbolic signals. Renewable energy is a highly significant negative predictor ($\beta=-0.307$, p = 0.003): a one-standard-deviation increase in renewable use is associated with a 0.31 standard-deviation reduction in the gap. Across the observed range (0 to 100 per cent) the implied reduction is roughly 0.88 standard deviations, about two-thirds of the sample standard deviation. Environmental capex is also significant and negative ($\beta=-0.224$, p = 0.044). Both findings support the signalling prediction that costly, physically verifiable commitments constrain a firm's ability to maintain a disclosure score misaligned with its operations (H1, H4).

\begin{table}[H]
\centering
\textbf{Table 7. Determinants of the Disclosure-Performance Gap, hybrid model (HC3 robust). Primary: full sample (n=200). Robustness: influence-trimmed (n=184).}\\[4pt]
\begin{tabular}{@{}
  >{\raggedright\arraybackslash}p{(\columnwidth - 12\tabcolsep) * \real{0.30}}
  >{\raggedright\arraybackslash}p{(\columnwidth - 12\tabcolsep) * \real{0.13}}
  >{\raggedright\arraybackslash}p{(\columnwidth - 12\tabcolsep) * \real{0.13}}
  >{\raggedright\arraybackslash}p{(\columnwidth - 12\tabcolsep) * \real{0.11}}
  >{\raggedright\arraybackslash}p{(\columnwidth - 12\tabcolsep) * \real{0.16}}
  >{\raggedright\arraybackslash}p{(\columnwidth - 12\tabcolsep) * \real{0.17}}@{}}
\toprule
\textbf{Variable} & \textbf{$\beta$ (n=200)} & \textbf{SE (HC3)} & \textbf{p} & \textbf{$\beta$ (n=184)} & \textbf{p (n=184)} \\
\midrule

Intercept & $-0.919$ & 0.354 & 0.010*** & $-0.509$ & 0.058* \\
Sector: Industrials (ref. Energy) & $-0.437$ & 0.248 & 0.078* & $-0.655$ & 0.000*** \\
Sector: Materials & $-0.405$ & 0.230 & 0.079* & $-0.549$ & 0.004*** \\
Sector: Utilities & $-0.291$ & 0.217 & 0.181 & $-0.400$ & 0.040** \\
\textbf{\% Renewable Energy (Std)} & $\mathbf{-0.307}$ & 0.103 & \textbf{0.003***} & $-0.301$ & 0.000*** \\
\textbf{Index Membership (0/1)} & $\mathbf{+0.782}$ & 0.175 & \textbf{0.000***} & +0.750 & 0.000*** \\
TCFD Supporter (0/1) & +0.858 & 0.349 & 0.014** & +0.521 & 0.075* \\
\textbf{\% Environmental Capex (Std)} & $\mathbf{-0.224}$ & 0.111 & \textbf{0.044**} & $-0.204$ & 0.007*** \\
\multicolumn{6}{@{}l@{}}{\footnotesize Primary (n=200): $R^2$=0.262, Adj.\ $R^2$=0.235, F(7,192)=9.74, p\textless0.001, AICc=617.24, BIC=642.88.} \\
\multicolumn{6}{@{}l@{}}{\footnotesize Robustness (n=184): $R^2$=0.341, Adj.\ $R^2$=0.315, F(7,176)=13.02, p\textless0.001, AICc=460.79, BIC=485.69.} \\
\bottomrule
\end{tabular}

\noindent\begin{minipage}{\linewidth}\footnotesize Note: *** p \textless{} 0.01, ** p \textless{} 0.05, * p \textless{} 0.10. Standard errors are HC3 robust. Dependent variable = DPG. Reference sector = Energy. AICc compares pools within a sample only and is not compared across the two samples. The TCFD coefficient is identified off 19 non-supporting firms and is read as directional rather than as a precise magnitude. Source: Author's own elaboration \end{minipage}\normalsize
\end{table}

Index membership is the single strongest predictor, associated with a gap wider by 0.782 standard deviations ($\beta=+0.782$, p \textless{} 0.001), after controlling for sector, renewable use, and capex. The stepwise algorithm retained index membership in preference to raw firm size, indicating that strategic market visibility, rather than asset scale, accounts for the association with decoupling. This supports H2 and is consistent with the coercive-isomorphism mechanism. TCFD support is positive and significant in the primary model ($\beta=+0.858$, p = 0.014), but it is identified off only 19 non-supporting firms and attenuates to marginal significance in the trimmed sample (p = 0.075), so it is read as a directional result rather than a reliable magnitude (H3). Importantly, the three other signals retain their sign and significance in the trimmed, raw-pool, and proxy-substituted specifications, so they do not depend on the inclusion of TCFD.

All three sector dummies are negative relative to Energy, and they are interpreted strictly as controls. Because the dependent variable is standardised within sector, the dummies absorb residual cross-sector scale differences after within-sector demeaning; they do not rank sectors by greenwashing and carry no independent greenwashing information. Finally, the exclusion of board diversity (H5), ROA (H6), SBTi validation (H7), and analyst coverage (H8) is a substantive result rather than a measurement failure: these variables were available and tested across pools, and their removal indicates that greenwashing is orthogonal to governance quality and financial resources once symbolic and substantive signals are controlled.

\subsection*{Diagnostic checks and assumption testing}

Having established the estimates, Table 8 reports the OLS assumption tests for the full ($n=200$) and trimmed ($n=184$) samples. In the full sample, residual normality fails (JB = 49.66, p \textless{} 0.001), driven by the 16 influential observations; after trimming, the cleaned sample passes (JB = 1.76, p = 0.414). Heteroscedasticity is present in both samples (Breusch-Pagan p = 0.038 and p = 0.018), consistent with error variance increasing with firm size under market-cap stratification, and is remediated by HC3 robust standard errors throughout. Multicollinearity is absent (maximum VIF 1.18). Linearity holds on visual inspection. The Durbin-Watson statistic is reported only as a formal artefact, since serial correlation is not a threat in a cross-section.

Taken together, the diagnostics confirm that the primary estimates support valid inference. The single failure, non-normal residuals in the full sample, is driven entirely by the 16 influential observations, does not bias the HC3 standard errors at this sample size, and disappears in the trimmed sample. Heteroscedasticity is handled by the HC3 correction throughout.

\begin{table}[H]
\centering
\textbf{Table 8. OLS assumption diagnostics, full and cleaned samples}\\[4pt]
\begin{tabular}{@{}
  >{\raggedright\arraybackslash}p{(\columnwidth - 8\tabcolsep) * \real{0.26}}
  >{\raggedright\arraybackslash}p{(\columnwidth - 8\tabcolsep) * \real{0.24}}
  >{\raggedright\arraybackslash}p{(\columnwidth - 8\tabcolsep) * \real{0.18}}
  >{\raggedright\arraybackslash}p{(\columnwidth - 8\tabcolsep) * \real{0.18}}
  >{\raggedright\arraybackslash}p{(\columnwidth - 8\tabcolsep) * \real{0.14}}@{}}
\toprule
\textbf{Assumption} & \textbf{Test} & \textbf{Full (n=200)} & \textbf{Clean (n=184)} & \textbf{Resolution} \\
\midrule

Normality of residuals & Jarque-Bera & JB=49.66, p\textless0.001 FAIL & JB=1.76, p=0.414 PASS & Outlier removal \\
Homoscedasticity & Breusch-Pagan & $\chi^2$=20.99, p=0.038 FAIL & $\chi^2$=16.88, p=0.018 FAIL & HC3 robust SE \\
Multicollinearity & VIF (max) & 1.18 PASS & 1.18 PASS & None required \\
Linearity & Residuals vs fitted & PASS (visual) & PASS (visual) & None required \\
Serial correlation & Durbin-Watson & 1.489 (artefact) & N/A, cross-section & N/A \\
\bottomrule
\end{tabular}

\noindent\begin{minipage}{\linewidth}\footnotesize Note: HC3 = heteroscedasticity-consistent standard errors (MacKinnon-White). The residual diagnostics before and after outlier removal, which show the normality recovery, are reported in Appendix A. Source: Author's own elaboration\end{minipage}\normalsize
\end{table}

\subsection*{Robustness and proxy sensitivity}

Five robustness checks confirm that the structural findings are not artefacts of the sample configuration, the choice of transformation, or the coding of the disclosure grade. Tables 9 and 10 report the headline coefficients across the comparable re-estimations of the primary model. Table 9 covers the influence-trimmed and firm-size-augmented specifications; Table 10 covers the exclusion of the thin TCFD variable and the rank-based reconstruction of the ordinal disclosure grade. A pure raw-variable specification (Pool A) preserves every directional finding and is reported in full in Table 11, since its unstandardised coefficients are not on the scale of the columns shown.

\begin{table}[!ht]
\centering
\textbf{Table 9. Robustness of the primary model (I): influence and specification}\\[4pt]
\begin{tabular}{@{}
  >{\raggedright\arraybackslash}p{(\columnwidth - 6\tabcolsep) * \real{0.40}}
  >{\centering\arraybackslash}p{(\columnwidth - 6\tabcolsep) * \real{0.20}}
  >{\centering\arraybackslash}p{(\columnwidth - 6\tabcolsep) * \real{0.20}}
  >{\centering\arraybackslash}p{(\columnwidth - 6\tabcolsep) * \real{0.20}}@{}}
\toprule
\textbf{Variable} & \textbf{Primary} & \textbf{Trimmed} & \textbf{+ Firm size} \\
 & (1) & (2) & (3) \\
\midrule
\% Renewable Energy (Std) & $-0.307$*** & $-0.301$*** & $-0.285$*** \\
Index Membership (0/1) & +0.782*** & +0.750*** & +0.689*** \\
TCFD Supporter (0/1) & +0.858** & +0.521* & +0.736** \\
\% Environmental Capex (Std) & $-0.224$** & $-0.204$*** & $-0.216$* \\
Firm size (Std) & n/a & n/a & +0.163* \\
\midrule
Adjusted $R^{2}$ & 0.235 & 0.315 & 0.245 \\
n & 200 & 184 & 200 \\
\bottomrule
\end{tabular}

\noindent\begin{minipage}{\linewidth}\footnotesize Note: HC3 robust standard errors throughout; all specifications include the sector dummies (Energy reference). *** p \textless{} 0.01, ** p \textless{} 0.05, * p \textless{} 0.10. Column (1) is the primary model of Table 7. Column (2) trims the 16 highest-influence observations by Cook's distance. Column (3) adds standardised firm size, which is significant only at the ten per cent level (p = 0.094) and does not absorb the index-membership effect. Source: Author's own elaboration \end{minipage}\normalsize
\end{table}

\begin{table}[!ht]
\centering
\textbf{Table 10. Robustness of the primary model (II): measurement of the disclosure variables}\\[4pt]
\begin{tabular}{@{}
  >{\raggedright\arraybackslash}p{(\columnwidth - 6\tabcolsep) * \real{0.40}}
  >{\centering\arraybackslash}p{(\columnwidth - 6\tabcolsep) * \real{0.20}}
  >{\centering\arraybackslash}p{(\columnwidth - 6\tabcolsep) * \real{0.20}}
  >{\centering\arraybackslash}p{(\columnwidth - 6\tabcolsep) * \real{0.20}}@{}}
\toprule
\textbf{Variable} & \textbf{Primary} & \textbf{Ex-TCFD} & \textbf{Rank gap} \\
 & (1) & (2) & (3) \\
\midrule
\% Renewable Energy (Std) & $-0.307$*** & $-0.353$*** & $-0.330$*** \\
Index Membership (0/1) & +0.782*** & +0.890*** & +0.971*** \\
TCFD Supporter (0/1) & +0.858** & n/a & +0.530* \\
\% Environmental Capex (Std) & $-0.224$** & $-0.199$* & $-0.235$** \\
\midrule
Adjusted $R^{2}$ & 0.235 & 0.204 & 0.252 \\
n & 200 & 200 & 199 \\
\bottomrule
\end{tabular}

\noindent\begin{minipage}{\linewidth}\footnotesize Note: HC3 robust standard errors throughout; all specifications include the sector dummies (Energy reference). *** p \textless{} 0.01, ** p \textless{} 0.05, * p \textless{} 0.10. Column (1) is the primary model of Table 7, repeated for comparison. Column (2) removes the TCFD indicator, identified off only 19 firms. Column (3) rebuilds the gap with its disclosure component re-standardised from within-sector ranks rather than the cardinal A-to-F coding; the rank-based and cardinal gaps correlate 0.96 (n = 199, one firm omitted for an indistinguishable gap value). Source: Author's own elaboration \end{minipage}\normalsize
\end{table}

Table 11 reports the raw-variable pool (Pool A) in full, providing the coefficients behind the directional check summarised in Tables 9 and 10.

\begin{table}[!ht]
\centering
\textbf{Table 11. Raw-variable specification (Pool A), HC3 robust, n=200}\\[4pt]
\begin{tabular}{@{}
  >{\raggedright\arraybackslash}p{(\columnwidth - 6\tabcolsep) * \real{0.46}}
  >{\centering\arraybackslash}p{(\columnwidth - 6\tabcolsep) * \real{0.18}}
  >{\centering\arraybackslash}p{(\columnwidth - 6\tabcolsep) * \real{0.16}}
  >{\centering\arraybackslash}p{(\columnwidth - 6\tabcolsep) * \real{0.20}}@{}}
\toprule
\textbf{Variable (raw scale)} & \textbf{Coefficient} & \textbf{p-value} & \textbf{Effect on gap} \\
\midrule
\% Renewable Energy (per point) & $-0.014$*** & \textless 0.001 & Narrows \\
Index Membership (0/1) & +0.835*** & \textless 0.001 & Widens \\
TCFD Supporter (0/1) & +1.040*** & 0.006 & Widens \\
\% Independent Directors (per point) & $-0.034$* & 0.057 & Narrows \\
\midrule
Adjusted $R^{2}$ & 0.228 & & \\
AICc & 619.12 & & \\
n & 200 & & \\
\bottomrule
\end{tabular}

\noindent\begin{minipage}{\linewidth}\footnotesize Note: Pool A is the best raw-only specification identified by the stepwise search, estimated with HC3 robust standard errors and sector dummies (Energy reference). Coefficients are on the raw scale and are therefore not comparable to the standardised models in Tables 7, 9, and 10: the renewable-energy and independent-director coefficients are per percentage point, while index membership and TCFD are zero-to-one effects. The three signals shared with the primary model keep their sign and significance, renewable energy narrowing the gap and index membership and TCFD widening it. Independent directors, a governance variable not selected by the hybrid model, enters this pool with a marginally significant negative coefficient. *** p \textless{} 0.01, * p \textless{} 0.10. Source: Author's own elaboration \end{minipage}\normalsize
\end{table}

\begin{figure}[H]
\centering
\caption{Renewable energy coefficient under the two disclosure proxies}\label{fig:renew}
\includegraphics[width=0.5\textwidth]{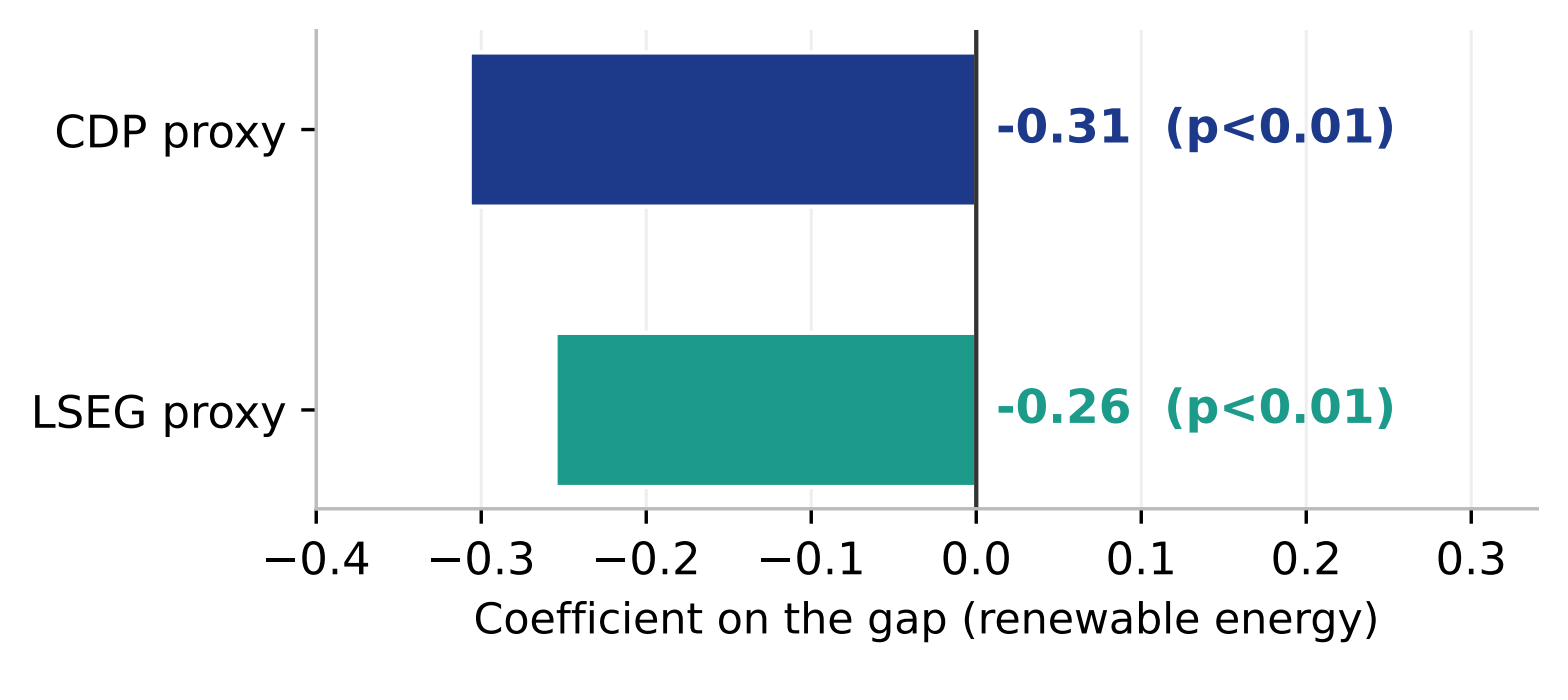}
\noindent\begin{minipage}{\linewidth}\footnotesize Note: Renewable energy coefficient under the two disclosure proxies. The negative effect is stable in sign and magnitude, $-0.31$ under the CDP gap and $-0.26$ under the LSEG gap, identifying it as the most robust finding in the study. Source: Author's own elaboration\end{minipage}\normalsize
\end{figure}

Figure~\ref{fig:renew} isolates the renewable-energy coefficient: it is the one effect that holds its sign and magnitude when the disclosure proxy is switched, from $-0.31$ under the CDP gap to $-0.26$ under the LSEG gap.

\begin{figure}[H]
\centering
\caption{Coefficient estimates under the CDP and LSEG disclosure proxies on the identical 200 firms}\label{fig:proxy}
\includegraphics[width=0.85\textwidth]{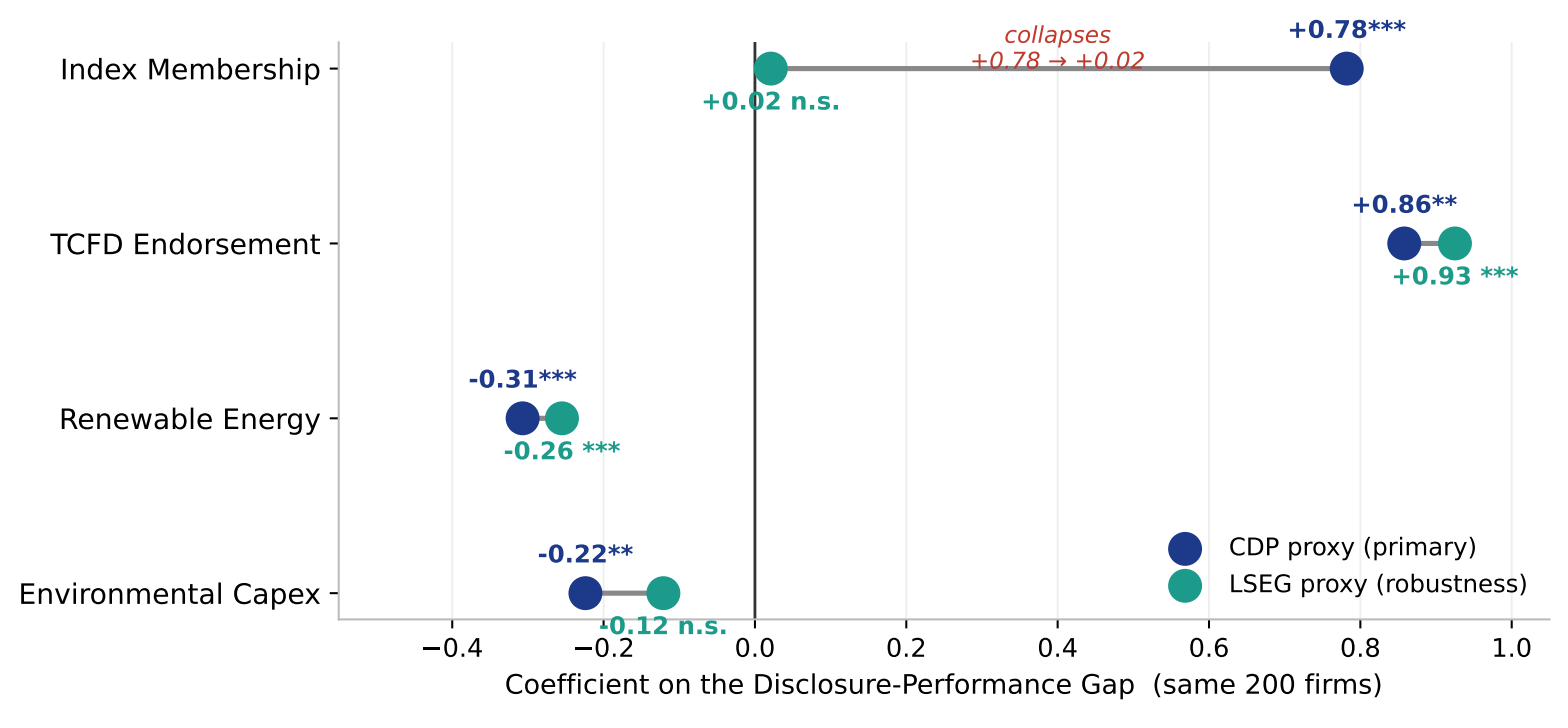}
\noindent\begin{minipage}{\linewidth}\footnotesize Note: The index-membership effect is the strongest predictor under CDP and collapses to zero under LSEG, while the renewable-energy effect survives both. Source: Author's own elaboration\end{minipage}\normalsize
\end{figure}

The most demanding check substitutes the disclosure proxy itself. The identical model is re-estimated on the identical 200 firms after replacing the CDP score with the LSEG Environmental Pillar Score, holding the Walk component and every other variable unchanged. This is a direct firm-level test of the aggregate confusion thesis: if two institutional disclosure proxies disagree, the determinants of the measured gap should not be stable across them. The two gap measures correlate at only 0.54 on the same firms. Table 12 reports the comparison, and Figure~\ref{fig:proxy} visualises the proxy-dependence of each coefficient.

\begin{table}[!ht]
\centering
\textbf{Table 12. Determinants of the gap under two disclosure proxies (standardised coefficients)}\\[4pt]
\begin{tabular}{@{}
  >{\raggedright\arraybackslash}p{(\columnwidth - 6\tabcolsep) * \real{0.34}}
  >{\raggedright\arraybackslash}p{(\columnwidth - 6\tabcolsep) * \real{0.22}}
  >{\raggedright\arraybackslash}p{(\columnwidth - 6\tabcolsep) * \real{0.22}}
  >{\raggedright\arraybackslash}p{(\columnwidth - 6\tabcolsep) * \real{0.22}}@{}}
\toprule
\textbf{Variable} & \textbf{CDP, n=200 (primary)} & \textbf{CDP, n=184 (trimmed)} & \textbf{LSEG, n=200} \\
\midrule

\% Renewable Energy (Std) & $-0.307$*** & $-0.301$*** & $-0.255$*** \\
Index Membership (0/1) & +0.782*** & +0.750*** & +0.021 (n.s.) \\
TCFD Supporter (0/1) & +0.858** & +0.521* & +0.925*** \\
\% Environmental Capex (Std) & $-0.224$** & $-0.204$*** & $-0.121$ (n.s.) \\
Adjusted $R^2$ & 0.235 & 0.315 & 0.087 \\
\bottomrule
\end{tabular}

\noindent\begin{minipage}{\linewidth}\footnotesize Note: standardised OLS coefficients with HC3 robust standard errors. The LSEG specification re-estimates the same model on the same 200 firms with the LSEG Environmental Pillar Score as the disclosure proxy. The two dependent variables correlate at 0.54. *** p \textless{} 0.01, ** p \textless{} 0.05, * p \textless{} 0.10; n.s. = not significant. Source: Author's own elaboration \end{minipage}\normalsize
\end{table}

Three patterns stand out. First, the renewable energy effect is negative and significant under both proxies ($-0.307$ under CDP, $-0.255$ under LSEG), the closest the data come to a proxy-independent signal of genuine decarbonisation. Second, the index-membership effect is not robust: it is the strongest predictor under CDP (+0.782, p \textless{} 0.001) but collapses to a precisely estimated zero under LSEG (+0.021, p = 0.91) on the very same firms. The greenwashing premium attached to index members is therefore a property of the CDP disclosure lens, not a stable property of the firms. Third, TCFD endorsement widens the gap under both proxies, while environmental capex is significant under CDP but not under LSEG, placing it among the proxy-fragile signals. Model fit also falls sharply, with adjusted $R^{2}$ dropping from 0.235 under CDP to 0.087 under LSEG. The substantive conclusion is that aggregate confusion is not only a feature of headline ratings but propagates into any single regression that uses one of those ratings as a disclosure proxy. A greenwashing finding is conditional on the rating lens applied, and this conditionality is the empirical core of our paper.

\section*{5.\quad CONCLUSIONS}

This paper provides evidence that the pre-CSRD voluntary reporting regime in European capital markets failed to align corporate environmental disclosure with operational performance. Rather than functioning as an accountability mechanism, the voluntary framework created a bifurcated market in which symbolic visibility was rewarded over substantive decarbonisation. This section interprets the central results, draws out the implications for investors and regulators, states the limitations of the design, and concludes.

The finding that flagship index membership is the strongest predictor of the gap, more powerful than firm size, sector identity, or any governance variable, is the study's most consequential result, and it must be read together with the proxy-sensitivity evidence. The index effect is large and highly significant under the CDP proxy but statistically indistinguishable from zero under the LSEG proxy on the identical firms. The defensible headline is therefore not that index membership universally produces greenwashing, but that the measured greenwashing of index members is highly conditional on the rating lens, which is precisely the aggregate confusion thesis applied at the firm level. Within the CDP lens, the mechanism is consistent with coercive isomorphism: passive investors use ESG scores in portfolio construction and proxy voting, creating a financial penalty for firms whose scores fall below investable thresholds and an incentive to optimise disclosure quality independent of operational progress. Meyer and Rowan's (1977) prediction is confirmed within that lens: the formal structures of environmental commitment are adopted and elaborated independently of their operational content. The practical corollary is that fund managers who use index inclusion as a positive ESG screen may, under the CDP lens, be selecting the highest-gap firms.

\newpage
The negative coefficients on renewable energy and environmental capex confirm the signalling separating equilibrium: firms that bear the operational and financial costs of genuine decarbonisation cannot simultaneously sustain a wide disclosure gap, because the physical and financial reality of their investments constrains the distance between claim and performance. A firm that commits large, irreversible capital to abatement does not need to inflate its disclosure score, because the asset base itself carries the signal. The management implication is that the most credible route to environmental credibility is operational investment, not communications.

The exclusion of board diversity, financial slack, SBTi validation, and analyst coverage from the model reframes greenwashing from a governance failure into a rational strategic choice. These four mechanisms are the primary non-regulatory tools through which the literature has theorised that greenwashing might be disciplined, and their collective failure to predict the gap implies that the current voluntary governance ecosystem, internal and external, does not constrain strategic decoupling. The irrelevance of profitability is particularly consequential: if greenwashing were a symptom of financial distress, ROA and the gap would be negatively related, yet profitable and struggling firms are statistically indistinguishable in their decoupling once symbolic-signal incentives are controlled. The irrelevance of SBTi validation is equally notable and may reflect both the use of ambitious targets as legitimacy shields and the 2023 SBTi pause for Oil and Gas, which removes the variable's variation from the Energy sector.

Three implications follow. First, mandatory disclosure alone is insufficient. The TCFD evidence shows that voluntary frameworks, even when widely adopted, do not close the gap; the value of the CSRD lies not in mandating more disclosure but in mandating audited performance metrics that cannot be optimised through narrative management. Second, the design of mandatory metrics matters: to break the decoupling equilibrium, the metrics must be auditable, standardised across sectors, and accompanied by financial consequences for non-performance, for example through integration with carbon pricing. Disclosure of emission intensity without financial consequences merely extends the voluntary era under a mandatory label. Third, institutional investors should shift their due diligence. The finding that index membership predicts a wider gap under the CDP lens implies that index-linked ESG integration carries an adverse selection problem, and that operational ratios, capex intensity and renewable share, are more reliable screening criteria for authentic transition risk than disclosure-based scores. The proxy-sensitivity result reinforces this: because the greenwashing a researcher or investor detects depends on the rating lens, the aggregate confusion that Berg, K\"olbel, and Rigobon (2022) document at the level of ratings reappears inside any single firm-level estimate. Until disclosure proxies are anchored to audited performance, conclusions about which firms are greenwashing remain conditional on the proxy chosen.

Several limitations bound these conclusions and define an agenda for future work. First, the dependent variable is restricted to Scope 1 and 2 emissions, creating a Scope 3 blind spot: for many Industrials and Materials firms, value-chain emissions are the dominant share of the footprint, so excluding them omits the part of the footprint where decoupling is easiest to hide. The direction of this bias is predictable: because firms can post strong Scope 1 and 2 figures while leaving large Scope 3 emissions undisclosed, the measured gap is most plausibly a lower bound on the true gap, and the estimates here are conservative. The CSRD mandates Scope 3 disclosure from 2025 to 2026, and the analysis should be replicated on the first cohort of audited Scope 3 data. Second, the cross-sectional design captures the year 2023 equilibrium but supports associational rather than causal inference; reverse causality and unobserved heterogeneity such as firm maturity or access to transition capital cannot be ruled out without panel data or a credible instrument, and longitudinal analysis across the CSRD transition years, ideally with firm fixed effects, will be essential to move toward causal claims. Third, the findings generalise to large-capitalisation, carbon-intensive European firms under a shared regulatory regime, and no further; they should not be extrapolated to small caps, to service sectors with negligible direct emissions, or to jurisdictions with different disclosure norms, and comparative work across U.S., Asian, and emerging markets is needed to test the external validity of the symbolic-substantive framework. Fourth, the proxy-sensitivity evidence rests on exactly two disclosure proxies, the CDP Climate Score and the LSEG Environmental Pillar Score. This pairing is sufficient to establish that a greenwashing finding is conditional on the rating lens, but two proxies cannot map the full distribution of disagreement across the rating industry, and the divergence observed here may understate or overstate the dispersion that a wider panel of raters would reveal. Extending the test to a third and fourth proxy, such as the S\&P Global Trucost or Sustainalytics environmental scores measured for the same firms in the same year, is the direct way to confirm that the conditionality generalises; the present result should be read as a firm-level demonstration of proxy dependence rather than as a complete characterisation of it. The deeper question of which proxy, if either, better corresponds to operational reality requires an external performance benchmark independent of every rating pipeline, and is left for subsequent work. Fifth, although the stepwise AICc search limits discretion and recovers the theory-specified signals, the candidate set still reflects theoretical priors, and signals not present in the 2023 data may become relevant as disclosure requirements expand.

The central contribution stands within these bounds. The case for the CSRD audit mandate does not rest on sector comparisons, which the within-sector construction cannot support. It rests on two firmer findings: the inability of the voluntary governance and verification ecosystem to constrain the gap, and the sensitivity of the measured gap to the choice of disclosure proxy. As long as Talk remains cheap and Walk remains expensive, firms in high-visibility institutional positions retain a rational incentive to decouple. The challenge for the post-CSRD era is to raise the cost of empty promises until the capital market rewards the Walk rather than the Talk.

\newpage
\section*{6.\quad REFERENCES}
\small
\setlength{\parskip}{4pt}
Becker, B. and Milbourn, T. (2011) \textquotesingle How did increased
competition affect credit ratings?\textquotesingle, Journal of Financial
Economics, 101(3), pp. 493--514.

Berg, F., Kölbel, J.F. and Rigobon, R. (2022) \textquotesingle Aggregate
confusion: The divergence of ESG ratings\textquotesingle, Review of
Finance, 26(6), pp. 1315--1344.

Bingler, J.A., Kraus, M., Leippold, M. and Webersinke, N. (2022)
\textquotesingle Cheap talk and cherry-picking: What ClimateBert has to
say on corporate climate disclosures\textquotesingle, SSRN Working Paper
No. 21-71.

Bloomberg Intelligence (2024) ESG assets rising to USD 40 trillion by
2030. Bloomberg. Available at:
\url{https://www.bloomberg.com/company/press/esg-assets-rising-to-40-trillion-by-2030-bloomberg-intelligence-finds/}
{[}Accessed: March 2026{]}.

Bourgeois, L.J. (1981) \textquotesingle On the measurement of
organizational slack\textquotesingle, Academy of Management Review,
6(1), pp. 29--39.

Breusch, T.S. and Pagan, A.R. (1979) \textquotesingle A simple test for
heteroscedasticity and random coefficient variation\textquotesingle,
Econometrica, 47(5), pp. 1287--1294.

Bromley, P. and Powell, W.W. (2012) \textquotesingle From smoke and
mirrors to walking the talk: Decoupling in the contemporary
world\textquotesingle, Academy of Management Annals, 6(1), pp. 483--530.

CDP (2023) CDP Climate Change Questionnaire 2023. London: Carbon
Disclosure Project.

Cho, C.H. and Patten, D.M. (2007) \textquotesingle The role of
environmental disclosures as tools of legitimacy: A research
note\textquotesingle, Accounting, Organizations and Society, 32(7--8),
pp. 639--647.

Cialdini, R.B. (1984) Influence: The Psychology of Persuasion. New
York: William Morrow.

Clarkson, P.M., Li, Y., Richardson, G.D. and Vasvari, F.P. (2008)
\textquotesingle Revisiting the relation between environmental
performance and environmental disclosure: An empirical
analysis\textquotesingle, Accounting, Organizations and Society,
33(4--5), pp. 303--327.

Connelly, B.L., Certo, S.T., Ireland, R.D. and Reutzel, C.R. (2011)
\textquotesingle Signaling theory: A review and
assessment\textquotesingle, Journal of Management, 37(1), pp. 39--67.

Cook, R.D. (1977) \textquotesingle Detection of influential observation
in linear regression\textquotesingle, Technometrics, 19(1), pp. 15--18.

Crawford, V.P. and Sobel, J. (1982) \textquotesingle Strategic
information transmission\textquotesingle, Econometrica, 50(6), pp.
1431--1451.

Delmas, M.A. and Burbano, V.C. (2011) \textquotesingle The drivers of
greenwashing\textquotesingle, California Management Review, 54(1), pp.
64--87.

DiMaggio, P.J. and Powell, W.W. (1983) \textquotesingle The iron cage
revisited: Institutional isomorphism and collective rationality in
organizational fields\textquotesingle, American Sociological Review,
48(2), pp. 147--160.

European Commission (2022) Directive (EU) 2022/2464 of the European
Parliament and of the Council --- Corporate Sustainability Reporting
Directive (CSRD). Official Journal of the European Union.

Hart, S.L. (1995) \textquotesingle A natural-resource-based view of the
firm\textquotesingle, Academy of Management Review, 20(4), pp.
986--1014.

Hurvich, C.M. and Tsai, C.L. (1989) \textquotesingle Regression and time
series model selection in small samples\textquotesingle, Biometrika,
76(2), pp. 297--307.

Ioannou, I. and Serafeim, G. (2015) \textquotesingle The impact of
corporate social responsibility on investment recommendations:
Analysts\textquotesingle{} perceptions and shifting institutional
logics\textquotesingle, Strategic Management Journal, 36(7), pp.
1053--1081.

Jensen, M.C. and Meckling, W.H. (1976) \textquotesingle Theory of the
firm: Managerial behavior, agency costs and ownership
structure\textquotesingle, Journal of Financial Economics, 3(4), pp.
305--360.

Joecks, J., Pull, K. and Vetter, K. (2013) \textquotesingle Gender
diversity in the boardroom and firm performance: What exactly
constitutes a "critical mass"?\textquotesingle, Journal of Business
Ethics, 118(1), pp. 61--72.

Kanter, R.M. (1977) Men and Women of the Corporation. New York: Basic
Books.

Kim, E.H. and Lyon, T.P. (2015) \textquotesingle Greenwash vs.
brownwash: Exaggeration and undue modesty in corporate sustainability
disclosure\textquotesingle, Organization Science, 26(3), pp. 705--723.

Kramer, V.W., Konrad, A.M. and Erkut, S. (2006) Critical Mass on
Corporate Boards: Why Three or More Women Enhance Governance. Wellesley
Centers for Women, Report No. WCW 11.

Lyon, T.P. and Maxwell, J.W. (2011) \textquotesingle Greenwash:
Corporate environmental disclosure under threat of
audit\textquotesingle, Journal of Economics and Management Strategy,
20(1), pp. 3--41.

Meyer, J.W. and Rowan, B. (1977) \textquotesingle Institutionalized
organizations: Formal structure as myth and ceremony\textquotesingle,
American Journal of Sociology, 83(2), pp. 340--363.

Science Based Targets initiative (SBTi) (2023) Companies Taking Action
Dashboard. Available at: https://sciencebasedtargets.org/ {[}Accessed:
November 2024{]}.

Spence, M. (1973) \textquotesingle Job market signaling\textquotesingle,
The Quarterly Journal of Economics, 87(3), pp. 355--374.

Spence, A.M. (1974) Market Signaling: Informational Transfer in Hiring
and Related Screening Processes. Cambridge, MA: Harvard University
Press.

STOXX (2023) STOXX® Europe 600 Index Guide. Zurich: STOXX Ltd.

Task Force on Climate-related Financial Disclosures (TCFD) (2023) 2023
Status Report. Basel: Financial Stability Board.

Waddock, S.A. and Graves, S.B. (1997) \textquotesingle The corporate
social performance-financial performance link\textquotesingle, Strategic
Management Journal, 18(4), pp. 303--319.

\normalsize

\newpage
\FloatBarrier
\section*{7.\quad APPENDICES}

\subsection*{Appendix A.\quad Regression diagnostics and coefficient robustness}

\begin{figure}[H]
\centering
\caption{Diagnostic plots for the full sample (n=200), before outlier removal}
\includegraphics[width=0.9\textwidth]{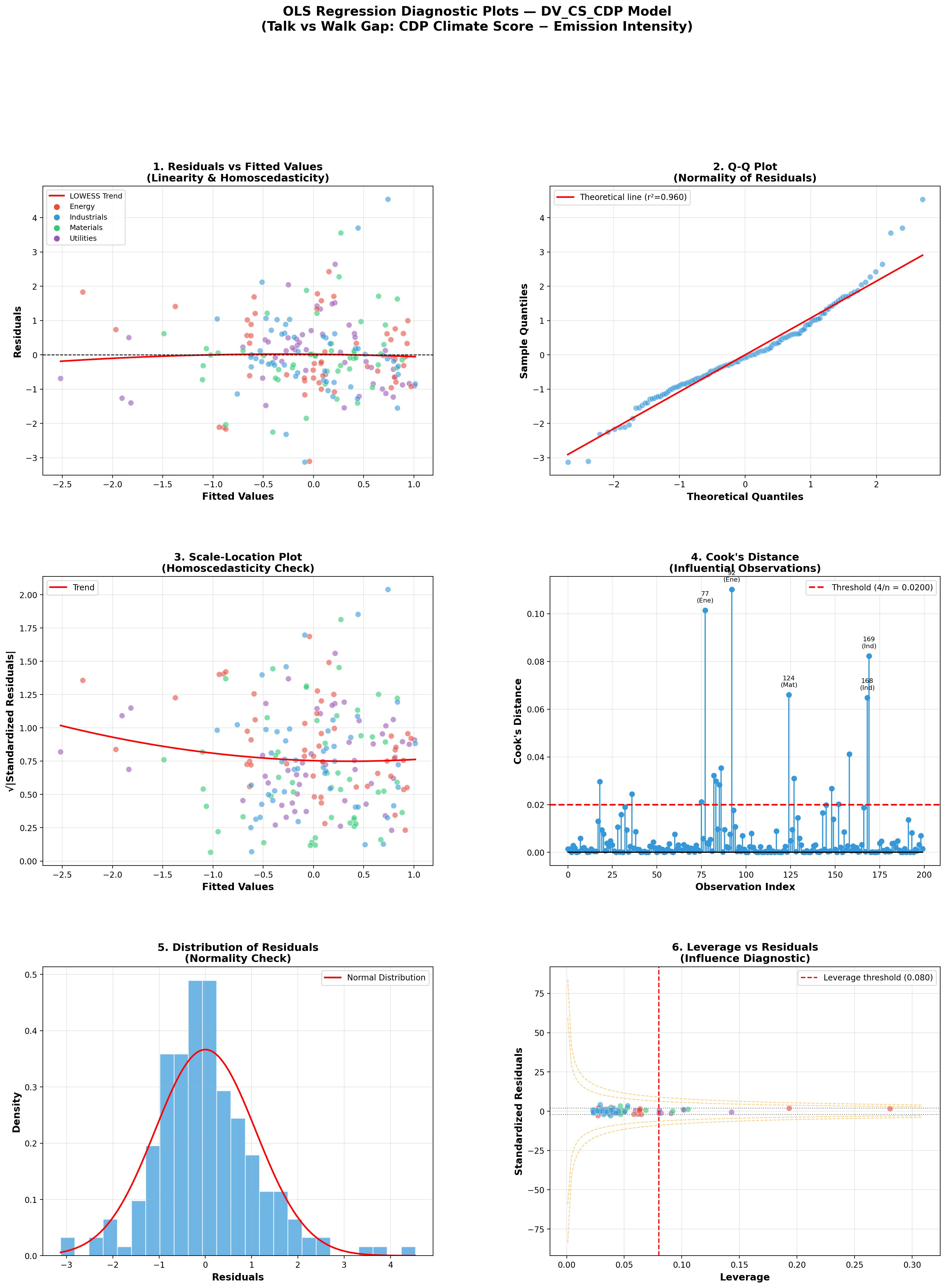}
\noindent\begin{minipage}{\linewidth}\footnotesize Note: The normal Q-Q plot and residual histogram show the departure from normality (Jarque-Bera = 49.66, p \textless{} 0.001) driven by the 16 influential observations. Source: Author's own elaboration \end{minipage}\normalsize
\end{figure}

\begin{figure}[H]
\centering
\caption{Diagnostic plots for the cleaned-sample specification (n=184), after outlier removal}
\includegraphics[width=0.9\textwidth]{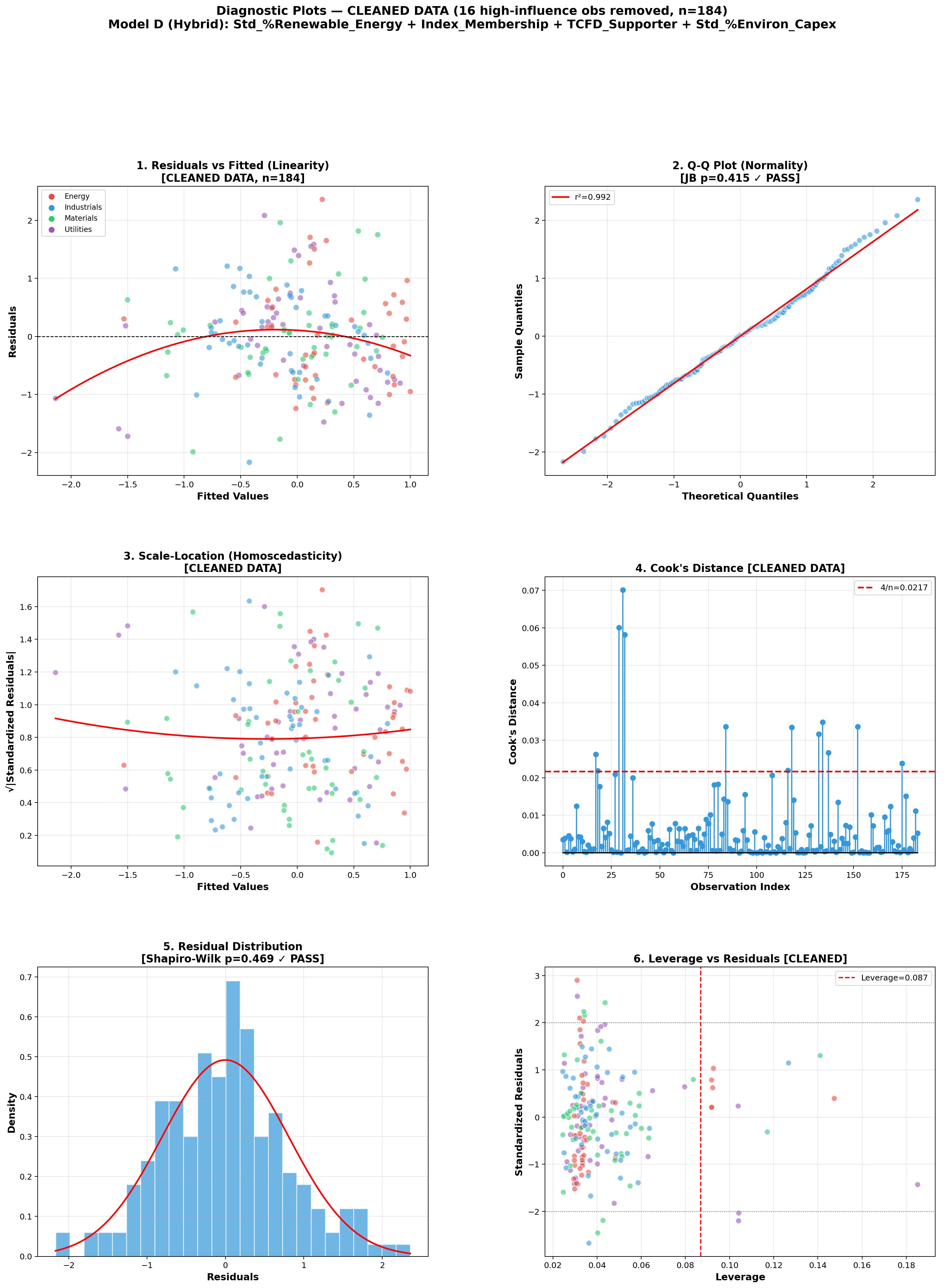}
\noindent\begin{minipage}{\linewidth}\footnotesize Note: Residual normality is restored (Jarque-Bera = 1.76, p = 0.414): residuals versus fitted, normal Q-Q, scale-location, and leverage. Source: Author's own elaboration \end{minipage}\normalsize
\end{figure}

\begin{figure}[H]
\centering
\caption{Coefficient estimates with HC3 95 per cent confidence intervals, full sample (n=200) versus cleaned sample (n=184)}
\includegraphics[width=0.98\textwidth]{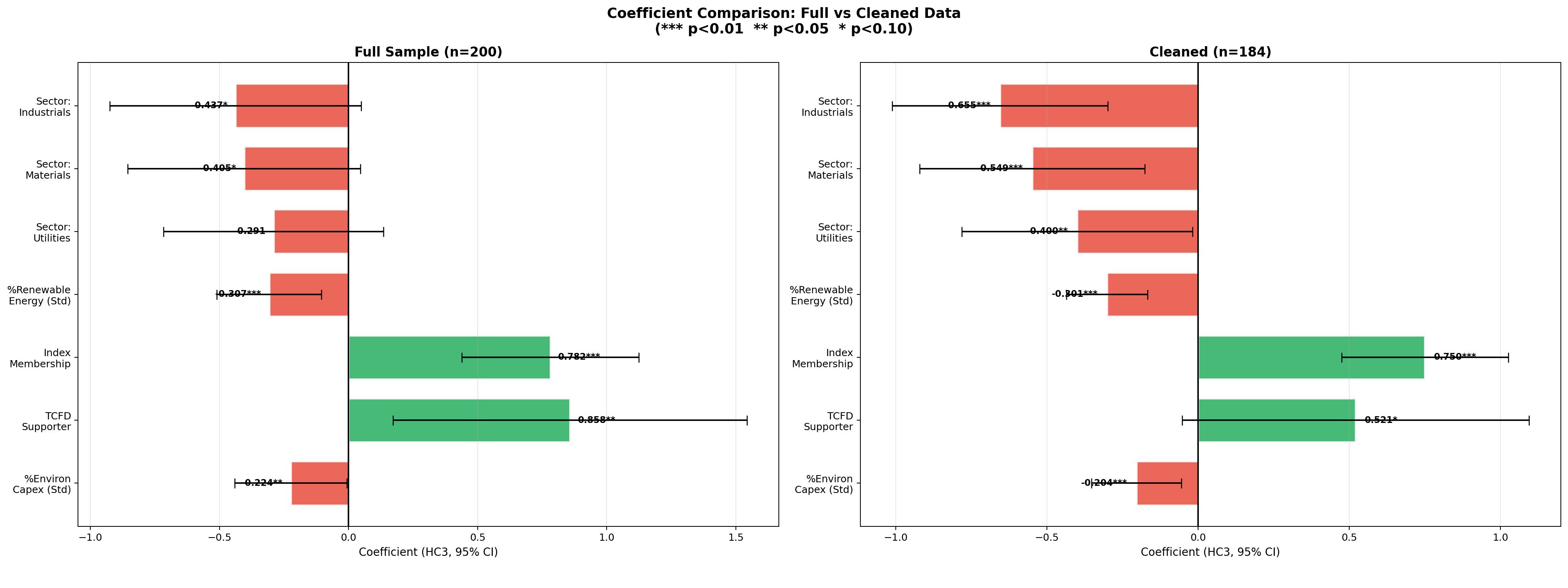}
\noindent\begin{minipage}{\linewidth}\footnotesize Note: Every structural predictor keeps its sign, magnitude, and significance across the two samples; the figure shows the confidence intervals that Table 7 reports only as standard errors. Source: Author's own elaboration \end{minipage}\normalsize
\end{figure}

\clearpage
\subsection*{Appendix B.\quad Correlation structure and pairwise relationships}

\begin{figure}[H]
\centering
\caption{Correlation matrix of the dependent variable and candidate predictors.}
\includegraphics[width=0.5\textwidth]{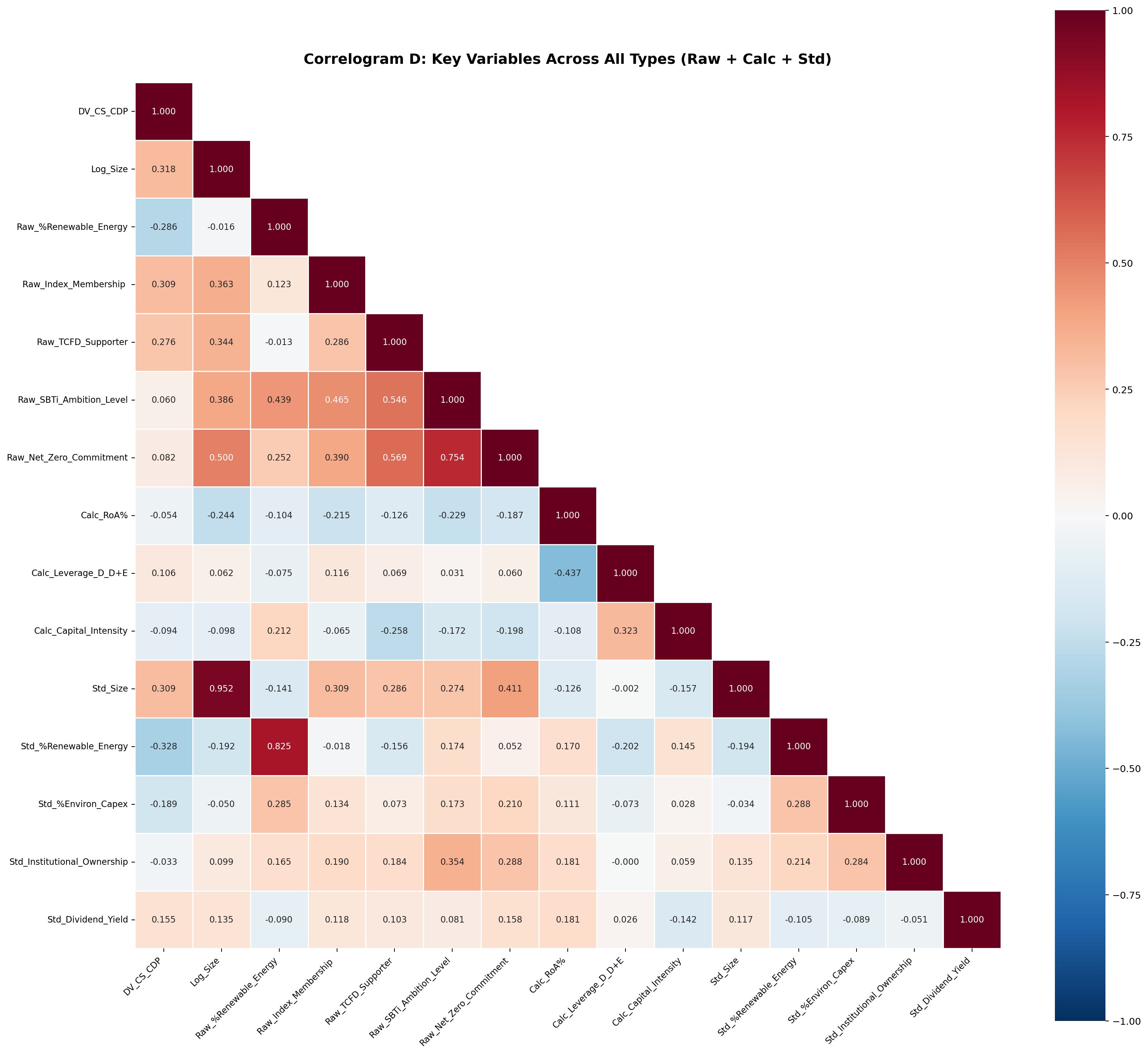}
\begin{center}
\footnotesize\textit{Source: Author's own elaboration}   
\end{center}
\end{figure}

\begin{figure}[H]
\centering
\caption{Pairwise relationships among the key variables.}
\includegraphics[width=0.5\textwidth]{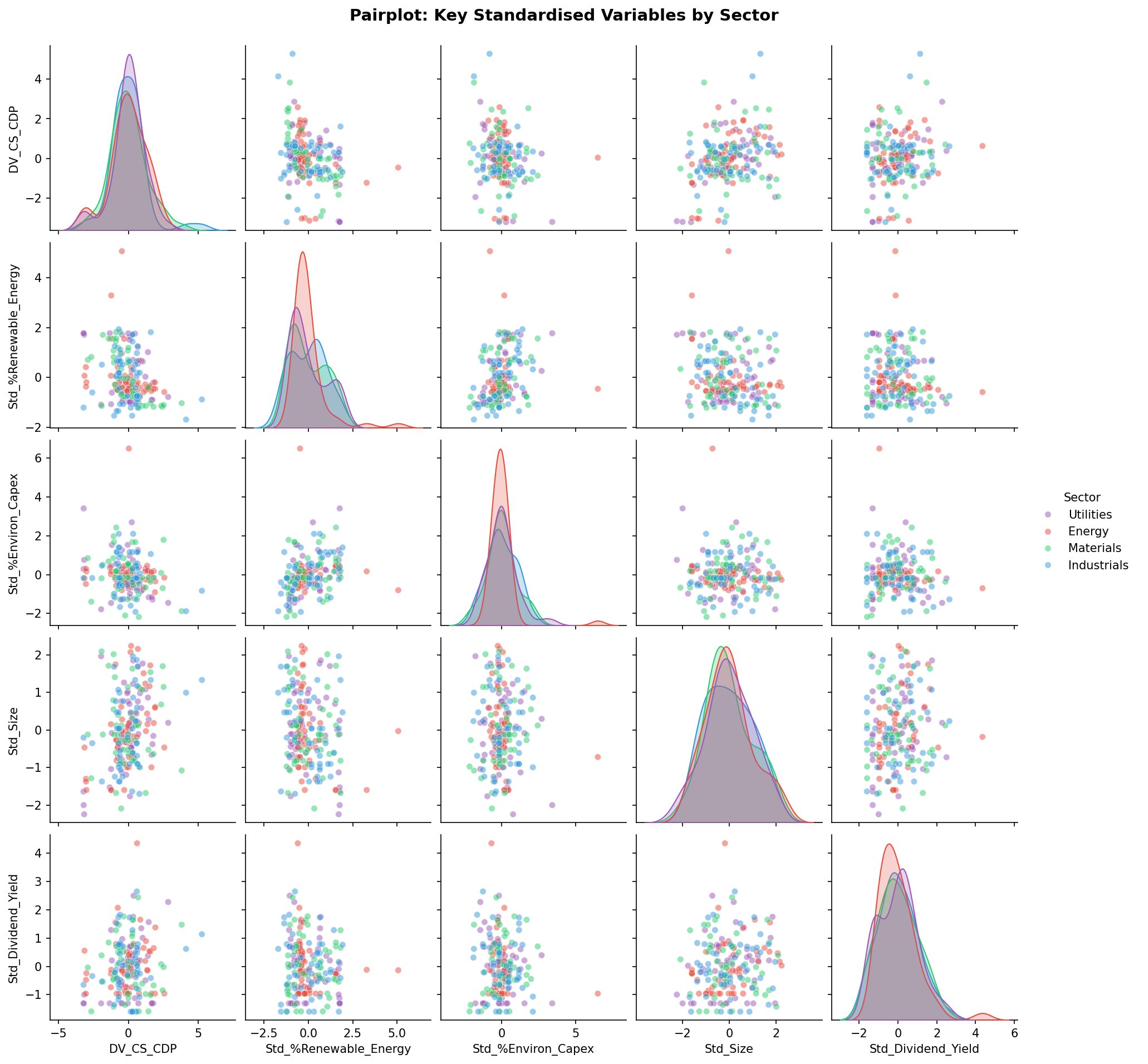}
\begin{center}
\footnotesize\textit{Source: Author's own elaboration}   
\end{center}
\end{figure}

\clearpage
\subsection*{Appendix C.\quad Candidate variable inventory}

The selection process in Section 3 began from 36 candidate variables. Table A1 lists them by scale type: raw values as reported, ratios calculated from raw inputs, and within-sample standardised transforms. Several variables appear in more than one form (for example, environmental capex as a raw share, a calculated ratio, and a standardised score); the multicollinearity filter in stage 3 retains only one form of each. The four variables selected into the final model are shown in bold.

\begin{table}[!ht]
\centering
\textbf{Table A1. The 36 candidate variables, by scale type}\\[4pt]
\begin{tabular}{@{}
  >{\raggedright\arraybackslash}p{(\columnwidth - 4\tabcolsep) * \real{0.16}}
  >{\raggedright\arraybackslash}p{(\columnwidth - 4\tabcolsep) * \real{0.84}}@{}}
\toprule
\textbf{Scale type} & \textbf{Variables} \\
\midrule

Raw (14) & Log size; log size centred by mean; current ratio; \% renewable energy; institutional ownership; \% independent directors; \% women on board; analyst coverage; cross-listing status; \textbf{index membership}; SBTi ambition level; years since SBTi commitment; \textbf{TCFD supporter}; net-zero commitment \\
Calculated (8) & Return on assets; return on equity; leverage (D/(D+E)); capital intensity; revenue growth; price-to-book ratio; dividend yield; \% environmental capex \\
Standardised (14) & Size; return on assets; return on equity; leverage; capital intensity; price-to-book ratio; revenue growth; dividend yield; institutional ownership; \% independent directors; \% women on board; \textbf{\% renewable energy}; environmental capex; \textbf{\% environmental capex} \\
\bottomrule
\end{tabular}

\noindent\begin{minipage}{\linewidth}\footnotesize Note: 14 raw, 8 calculated, and 14 standardised forms, 36 variables in total. Bold entries are the four predictors retained in the final model (Section 3, equation 5). Three sector dummies (Industrials, Materials, Utilities; Energy as reference) are added as controls. Source: Author's own elaboration \end{minipage}\normalsize
\end{table}

\clearpage
\subsection*{Appendix D.\quad Consolidated and sector-specific model equations}

The consolidated specification, with Energy as the reference sector, is
\begin{equation*}
\mathrm{DPG}_{i} = \beta_{0} + \beta_{1}\mathrm{Renew}^{\mathrm{Std}}_{i} + \beta_{2}\mathrm{Index}_{i} + \beta_{3}\mathrm{TCFD}_{i} + \beta_{4}\mathrm{Capex}^{\mathrm{Std}}_{i} + \gamma_{1}\mathrm{Ind}_{i} + \gamma_{2}\mathrm{Mat}_{i} + \gamma_{3}\mathrm{Util}_{i} + \varepsilon_{i}.
\end{equation*}
Setting each sector indicator to its value yields four sector-specific equations with common slopes and a sector-shifted intercept. Substituting the primary full-sample estimates ($\beta_{0}=-0.919$, $\beta_{1}=-0.307$, $\beta_{2}=+0.782$, $\beta_{3}=+0.858$, $\beta_{4}=-0.224$, $\gamma_{1}=-0.437$, $\gamma_{2}=-0.405$, $\gamma_{3}=-0.291$) gives:
\begin{align*}
\text{Energy:}\quad & \mathrm{DPG}_{i} = -0.919 -0.307\,\mathrm{Renew}^{\mathrm{Std}}_{i} +0.782\,\mathrm{Index}_{i} +0.858\,\mathrm{TCFD}_{i} -0.224\,\mathrm{Capex}^{\mathrm{Std}}_{i} + \varepsilon_{i} \\
\text{Industrials:}\quad & \mathrm{DPG}_{i} = -1.356 -0.307\,\mathrm{Renew}^{\mathrm{Std}}_{i} +0.782\,\mathrm{Index}_{i} +0.858\,\mathrm{TCFD}_{i} -0.224\,\mathrm{Capex}^{\mathrm{Std}}_{i} + \varepsilon_{i} \\
\text{Materials:}\quad & \mathrm{DPG}_{i} = -1.324 -0.307\,\mathrm{Renew}^{\mathrm{Std}}_{i} +0.782\,\mathrm{Index}_{i} +0.858\,\mathrm{TCFD}_{i} -0.224\,\mathrm{Capex}^{\mathrm{Std}}_{i} + \varepsilon_{i} \\
\text{Utilities:}\quad & \mathrm{DPG}_{i} = -1.210 -0.307\,\mathrm{Renew}^{\mathrm{Std}}_{i} +0.782\,\mathrm{Index}_{i} +0.858\,\mathrm{TCFD}_{i} -0.224\,\mathrm{Capex}^{\mathrm{Std}}_{i} + \varepsilon_{i}
\end{align*}
The slopes are identical across sectors by construction; only the intercept shifts. Because the gap is standardised within sector, these intercept shifts are mechanical controls for residual scale differences and are not interpreted as a ranking of greenwashing across sectors.

\end{document}